\documentstyle[epsfig,aps,floats,eqsecnum,preprint]{revtex}

\def\laq{\raise 0.4ex\hbox{$<$}\kern -0.8em\lower 0.62
ex\hbox{$\sim$}}
\def\gaq{\raise 0.4ex\hbox{$>$}\kern -0.7em\lower 0.62ex\hbox{$\sim$}}

\font\tenbb=msbm10
\font\sevenbb=msbm7
\font\fivebb=msbm5
\newfam\bbfam
\textfont\bbfam=\tenbb \scriptfont\bbfam=\sevenbb
\scriptscriptfont\bbfam=\fivebb

\def\bb{\fam\bbfam}
\def\Rb{{\bb R}}

\newcommand{\vrsmall}{\vrule width 0pt height 18pt depth 15pt}

\newcommand{\pa}{\partial} 

\newcommand{\vphi}{\varphi}

\newcommand{\beq}{\begin{equation}} 
\newcommand{\eeq}{\end{equation}}
\newcommand{\bea}{\begin{eqnarray}} 
\newcommand{\eea}{\end{eqnarray}}
\newcommand{\beam}{\begin{mathletters}} 
\newcommand{\eeam}{\end{mathletters}}
\newcommand{\var}{\rm Var} 
\newcommand{\ophi}{\overline{\phi}} 
\newcommand{\obeta}{\overline{\beta}} 
\newcommand{\om}{\overline{m}}

\begin{document}
\draft
\preprint{\vbox{\baselineskip=12pt
\rightline{IHES/P/98/44} 
\vskip 0.2truecm
\rightline{CERN-TH/98-187}
\vskip 0.2truecm
\rightline{hep-th/9806230}}}

\title{\Large\bf Pre-big bang bubbles from the gravitational instability of 
generic string vacua }
\author{A. Buonanno${}^{a}$, T. Damour${}^{(a,b)}$,
G. Veneziano${}^{(c,a)}$}
\address{$^a$ {\it Institut des Hautes Etudes Scientifiques, 91440
Bures-sur-Yvette, France} \\ 
{$^b$ {\it DARC, CNRS-Observatoire de Paris, 92195 Meudon, France}} \\
{$^c$ {\it Theory Division, CERN, CH-1211 Geneva 23, Switzerland}}}
\maketitle
\vskip -0.8truecm
\begin{abstract}
We formulate the basic postulate of pre-big bang cosmology 
as one of ``asymptotic past triviality", by which we mean 
that the initial state is a generic perturbative solution 
of the tree-level low-energy effective action. Such a 
past-trivial ``string vacuum'' is made of an arbitrary 
ensemble of incoming gravitational and dilatonic waves, and 
is generically prone to gravitational instability, leading 
to the possible formation of many black holes hiding singular 
space-like hypersurfaces. Each such singular space-like 
hypersurface of gravitational collapse becomes, in the 
string-frame metric, the usual big-bang $t=0$ hypersurface, 
i.e. the place of birth of a baby Friedmann universe after a 
period of dilaton-driven inflation. Specializing to the 
spherically-symmetric case, we review and reinterpret previous 
work on the subject, and propose a simple, scale-invariant 
criterion for collapse/inflation in terms of asymptotic data 
at past null infinity. Those data should determine whether, 
when, and where collapse/inflation occurs, and, when it does, 
fix its characteristics, including  anisotropies on the big bang 
hypersurface whose imprint could have survived till now.
Using Bayesian probability concepts, we finally attempt to 
answer some fine-tuning objections recently moved to the pre-big 
bang scenario.
\end{abstract}
\maketitle

\section{Introduction and general overview}
\label{sec1}
Superstring theory (see \cite{GSW} for a review) is the only presently known framework 
in which gravity can be consistently quantized, at least perturbatively.
The  well-known difficulties met in
trying to quantize General Relativity (GR) --or its supersymmetric extensions-- are
avoided, in string theory, by the presence  of
 a fundamental quantum of length \cite{stringlength}  $\ell_s \sim  (\alpha')^{1/2}$. 
Thus, at distances
shorter than $\ell_s$, 
string gravity is expected to be drastically different from --and in particular
to be much ``softer" than--  General Relativity.

However, as was noticed since the early days of string theory \cite{joel}, 
a conspicuous difference between string and Einstein gravity persists even at
low energies (large distances).
Indeed, a striking prediction of string theory is that its ``gravitational sector'' is 
richer than that of GR: in particular, all versions of string 
theory predict the existence of a scalar partner of the spin-two  
graviton, i.e. of the metric tensor $g_{\mu \nu}$, the so-called dilaton, $\phi$.
This field plays a central r\^ole 
within string theory \cite{GSW} since its present vacuum expectation value (VEV) $\phi_{\rm now}$ 
fixes the string coupling constant $g^2 = e^{\phi_{\rm now}}$, and, through
it, the present values of gauge and gravitational couplings. In particular, it fixes the ratio
of $\ell_s$ to the Planck length $\ell_P \sim G^{1/2}$ by $\ell_P \sim g\,\ell_s$. 
The relation is such that   
gauge and gravitational couplings unify at the string energy scale, i.e.
 at  $E \sim M_s \sim g M_P \sim (\alpha')^{-1/2} \sim 3 
\times 10^{17}$ GeV. It thus seems that the string way to quantizing gravity
forces this new particle/field upon us.

We believe that the dilaton represents an interesting 
prediction (an opportunity rather than a nuisance) whose possible existence 
should be taken seriously, and whose observable consequences should be 
carefully studied. 
Of course, tests of GR \cite{Damour} put severe constraints on what
the dilaton can do today. The simplest way to recover GR at late times
is to assume \cite{TV} that $\phi$ gets a mass from supersymmetry-breaking non-perturbative 
effects. Another possibility might be 
to use the string-loop modifications of the dilaton couplings for driving 
$\phi$ toward a special value where it decouples from matter \cite{DP}. 
These alternatives do not rule out the possibility
that the dilaton may have had an important r\^ole  in the previous  history of the
universe. Early cosmology stands out as a particularly 
interesting arena where to study both the dynamical effects of the dilaton
and those associated with the existence of a fundamental length in string theory.

In a series of previous papers \cite{1}, \cite{PBB} a model of early string cosmology, in 
which the dilaton plays a key dynamical r\^ole, was introduced and developed: the so-called 
pre-big bang (PBB) scenario. One of the key ideas of this scenario is to use 
the kinetic energy of the dilaton to drive a period of inflation of the 
universe. The motivation is that the presence of a (tree-level 
coupled) dilaton essentially destroys\cite{BS} the usual inflationary mechanism 
\cite{Linde}: instead of driving an exponential inflationary expansion, a 
(nearly) constant vacuum energy drives the string coupling $g = e^{\phi / 2}$ 
towards small values, thereby causing the universe to expand only as a small 
power of time. If one takes seriously the existence of the dilaton, the PBB 
idea of a dilaton-driven inflation offers itself as one of the very few natural 
ways of reconciling string theory and inflation. Actually, the existence 
of inflationary solution in string cosmology is a consequence of its 
(T-) duality symmetries \cite{2}. 

This paper develops further the PBB scenario by presenting a very general class 
of possible initial states for string cosmology, and by describing their 
subsequent evolution, via gravitational instability, into a multi-universe 
comprising (hopefully) sub-universes looking like ours. This picture generalizes, and makes 
more concrete, recent work \cite{V97}, \cite{spanish} about inhomogeneous 
versions of pre-Big Bang cosmology.

Let us first recall that an inflation driven by the kinetic 
energy of $\phi$ forces both the coupling and 
the curvature to {\it grow} during inflation \cite{1}. This implies that the 
initial state must be very perturbative in two respects: i) it must have
very small initial curvatures (derivatives) in string units and, ii) it must exhibit 
a tiny initial coupling 
 $g_i = e^{\phi_i /2} \ll 1$.  As the string 
coupling $g^2$ measures the strength of quantum corrections (i.e. plays the 
r\^ole of $\hbar$ in dividing the lagrangian: ${\cal L} = g^{-2} \, {\cal 
L}'$), quantum (string-loop) corrections 
are initially negligible. Because of i), $\alpha^\prime$
corrections can also be neglected. In conclusion, dilaton-driven inflation must start
from a regime in which the tree-level low-energy approximation 
to string theory is extremely accurate, something we may call an asymptotically trivial
state.

In the present paper, 
we consider a very general class of such ``past-trivial'' states. Actually, 
perturbative string theory is well-defined only when one considers such classical 
states as background, or ``vacuum'', configurations. For the sake
of simplicity the set of string 
vacua that we consider are already compactified to
 four dimensions and are truncated to the 
gravi-dilaton sector (antisymmetric tensor and moduli being set to zero). 
Within these limitations, the set of all perturbative string vacua coincides 
with the generic solutions of the tree-level low-energy effective action 
\cite{Fradkin} 
\beq
S_s = \frac{1}{\alpha^\prime} \int d^4 x \, \sqrt{G} \, e^{-\phi} [R(G) + G^{\mu 
\nu} \partial_{\mu} \phi \, \partial_{\nu} \phi] \,,
\label{eqn1.1}
\eeq
where we have denoted by $G_{\mu \nu}$ the string-frame ($\sigma$-model) metric.
The generic solution is parametrized by 6 functions of three variables.  These 
functions can be thought of classically as describing the two helicity$-2$ 
modes of gravitational waves, plus the helicity$-0$ mode of dilatonic waves. 
[Each mode being described by two real functions corresponding, e.g., to the 
Cauchy data $(\phi , \dot \phi)$ at some initial time.] The same counting of the 
degrees of freedom in selecting string vacua can  be obtained by considering all the 
marginal operators (i.e. all conformal-invariance-preserving 
continuous deformations) of the conformal field 
theory defining the quantized string in trivial space-time.
 We therefore envisage, as initial state, 
the most general past-trivial classical solution of (\ref{eqn1.1}), i.e. an 
arbitrary ensemble of incoming gravitational and dilatonic waves. 

\begin{figure}
\centerline{\epsfig{file=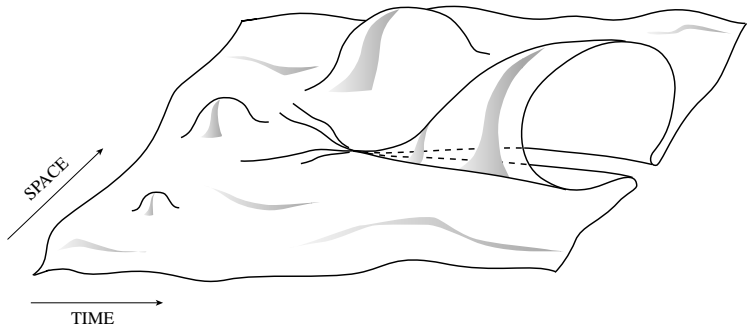,width=0.6\textwidth,angle=0}}
\vskip 0.5truecm
\caption{\sl Symbolic sketch of the birth of many pre-big bang 
bubbles from the gravitational instability of a generic string vacuum 
made of a stochastic bath of classical incoming gravitational 
and dilatonic waves. Each local Einstein-frame collapse 
of sufficiently strong waves forms a cosmological-like 
space-like singularity hidden behind a black hole. The parts of those classical
singularities where the string coupling grows inflate, when viewed 
in the string frame, and generate ballooning patches of space 
(here schematized as the stretching of one spatial dimension) 
which are expected to evolve into many separate quasi-closed Friedmann 
hot universes.}
\label{fig1}
\end{figure}

\begin{figure}
\centerline{\epsfig{file=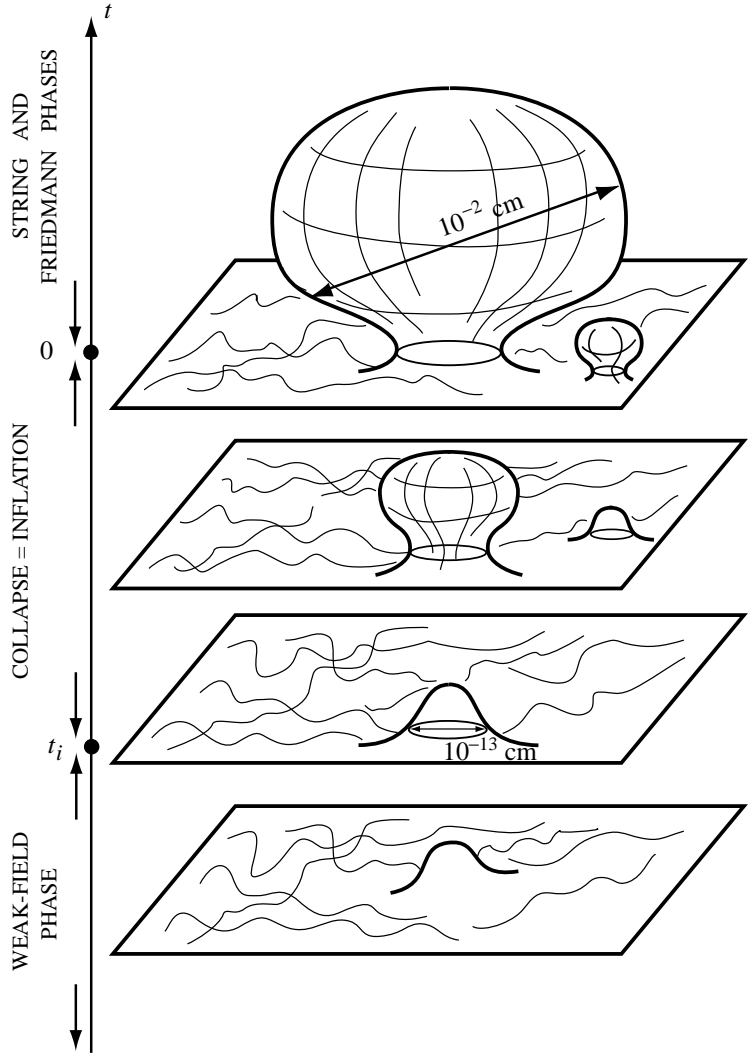,width=0.6\textwidth,angle=0}}
\vskip 0.5truecm
\caption{\sl A representation of pre-big bang bubbles similar 
to that of Fig.~\ref{fig1}, but in 2+1 dimensions. The different
horizontal planes represent different instants in the evolution 
from the asymptotic trivial past to the 
Friedmann phase. Two inflationary bubbles characterized by two different 
initial horizon sizes (both large in string units) are shown
to lead to Universes of very different homogeneity scale 
at the time at which the Hubble radius reaches string-scale values 
($\ell_s = {\cal O}(10^{-32}\, {\rm cm})$).}
\label{fig2}
\end{figure}

Our aim will be to show how such a stochastic bath of classical 
incoming waves (devoid of any ordinary matter) can evolve into our rich,
complex, expanding universe. The basic mechanism we consider for turning such a 
trivial, inhomogeneous and anisotropic, initial state into a Friedmann-like 
cosmological universe is gravitational instability (and quantum particle production as far as
heating up the universe is concerned \cite{PBB}). We find that, when the 
initial waves satisfy a certain (dimensionless) strength criterion, they 
collapse (when viewed in the Einstein conformal frame) under their own weight. 
When viewed in the (physically most appropriate) string conformal frame, each 
gravitational collapse leads to the local birth of a baby inflationary universe 
blistering off the initial vacuum. Thanks to the peculiar properties of 
dilaton-driven inflation (i.e. the peculiar properties of collapse with the 
equation of state $p = \epsilon$), each baby universe is found to contain a 
large homogeneous patch of expanding space which might constitute the beginning 
of a local Big Bang. We then expect each of these ballooning patches 
of space to evolve into a quasi-closed Friedmann universe
\footnote{Our picture of baby universes created by gravitational 
collapse is reminiscent of earlier proposals \cite{FMM}, \cite{MB}, 
\cite{Smolin}, 
but differs from them by our crucial use of string-theory motivated ideas.}.
This picture is sketched in Fig.~\ref{fig1} and Fig.~\ref{fig2}.
In order to study in detail this scenario, we focus, in this 
paper, on the technically simplest case containing nontrivial incoming waves 
able to exhibit gravitational instability: spherically symmetric dilatonic 
waves. This simple toy model seems to contain many of the key physical 
features we wish to study.

Before entering the technicalities of this model, let us clarify a few general 
methodological issues. One of the goals of theoretical cosmology is to 
``explain'' the surprisingly rich and special structure of our universe. 
However, the concept of ``explanation'' is necessarily dependent on one's 
prejudices and taste. We wish only to show here how,  modulo an 
``exit'' assumption \cite{PBB}, 
one can ``explain'' the appearance of a hot, expanding 
homogeneous universe starting from a generic classical inhomogeneous vacuum of 
string theory. We do not claim that this scenario solves all the cosmological 
``problems'' that one may wish to address (e.g., we leave untouched here the 
monopole and gravitini problems). We do not try either to ``explain'' the 
appearance of our universe as a quantum fluctuation out of ``nothing''. We 
content ourselves by assuming, as is standard in perturbative string theory, 
the existence of a classical vacuum and showing how gravitational instability 
can then generate some interesting qualitatively new structures akin to those 
of our universe. In particular, we find it appealing to ``understand'' the striking 
existence of a preferred rest frame in our universe, starting from a 
stochastic bath of waves propagating with the velocity of light and thereby 
exhibiting no clear split between space and time. We shall also address in 
this paper the question of the naturalness of our scenario. Recent papers 
\cite{TW}, \cite{KLB} have insisted on the presence of two large numbers among 
the parameters defining the PBB scenario. Our answer to this issue (which is, 
anayway, an issue of taste and not a scientifically well posed problem) is 
twofold: on the one hand, we point out that the two large numbers in question 
are classically undefined and therefore irrelevant when discussing the 
naturalness of a {\it classical} vacuum state; on the other hand, we show 
that the selection effects 
associated to asking any such ``fine-tuning'' question can render natural the 
presence of these very large numbers.

\section{Asymptotic null data for past-trivial string vacua}
\label{sec2}
Motivated by the pre-Big Bang idea of an initial weak-coupling, low-curvature 
state we consider the tree-level (order $g^{-2}$) effective action for the 
gravitational sector of critical superstring theory, taken at lowest order in 
$\alpha'$, Eq.~(\ref{eqn1.1}). We set to zero the antisymmetric tensor field 
$B_{\mu \nu}$, and work directly in 4 dimensions (i.e., we assume that the 
gravitational moduli describing the 6 ``internal dimensions'' are frozen). 
Though the physical interpretation of our work is best made in terms of the 
original string (or $\sigma$-model) metric $G_{\mu \nu}$ appearing in 
Eq.~(\ref{eqn1.1}), it will be technically convenient to work with the 
conformally related Einstein metric
\beq
g_{\mu \nu} = e^{-(\phi -\phi_{\rm now})} \, G_{\mu \nu} \, , 
\quad \quad 16 \pi G = \alpha^\prime \, e^{\phi_{\rm now}}\,. 
\label{eqn2.1}
\eeq
In the following, we will set $G = 1$.
In terms of the Einstein metric $g_{\mu \nu}$, the low-energy tree-level string 
effective action (\ref{eqn1.1}) reads
\footnote{We will use the signature $(-,+,+,+)$ and the conventions
${R}^\mu_{\,\,\,\,\nu \rho \sigma} = \pa_\rho \Gamma^\mu_{\nu\sigma}
- \dots\,,\,\,{R}_{\mu \nu} = {R}^\rho_{\,\,\,\,\mu \rho \nu}$.}
\beq
\label{1.1}
S = \frac{1}{16 \pi}\int d^4 x \,\sqrt{g}\,
\left [ {R} - \frac{1}{2}\pa_\mu \phi\, \pa^\mu \phi \right ]\,.
\eeq
The corresponding classical field equations are
\beq
R_{\mu \nu} = \frac{1}{2} \,\partial_{\mu} \phi \, \partial_{\nu} \phi \, ,  
\label{eq1.4} 
\eeq
\beq
\Box \, \phi \equiv \nabla^{\mu} \, \nabla_{\mu} \, \phi = 0 \, ,
\label{eqn2.3}
\eeq
with Eq. (\ref{eqn2.3}) actually following 
from Eq.~(\ref{eq1.4}) thanks to the Bianchi identity.
As explained in the Introduction, we consider a generic solution of these 
classical field equations admitting an asymptotically trivial past. Such 
an asymptotic ``incoming'' classical state should allow a description in terms 
of a superposition of ingoing gravitational and dilatonic waves. The 
work of Bondi, Sachs, Penrose \cite{BSP} and many others in classical gravitation theory 
indicates that this incoming state should exhibit a regular past null infinity 
${\cal I}^-$, and should be parametrizable by some asymptotic null data (i.e. 
conformally renormalized data on ${\cal I}^-$). In plain terms, this means the 
existence of suitable asymptotically Minkowskian \footnote{These coordinates 
must be restricted by the condition that the incoming coordinate cones $v 
\equiv t + r = {\rm const.}$ 
be asymptotically tangent to exact cones of $g_{\mu 
\nu} (x^{\lambda})$.} coordinates $(x^{\mu}) = (t,x,y,z)$ (which can then be 
used in the standard way to define polar coordinates $r,\theta,\varphi$ and the 
advanced time $v \equiv t + r \equiv t + (x^2 + y^2 + z^2)^{1/2}$) such that 
the following expansions hold when $r \rightarrow \infty$, at fixed $v$, 
$\theta$ and $\varphi$:
\beq
\phi (x^{\lambda}) = \phi_0 + \frac{f(v,\theta , \varphi)}{r} + {o} \left( 
\frac{1}{r} \right) \, , 
\label{eqn2.4}
\eeq
\beq
g_{\mu \nu} (x^{\lambda}) = \eta_{\mu \nu} + \frac{f_{\mu \nu} (v,\theta , 
\varphi)}{r} + {o} \left( \frac{1}{r} \right) \, . 
\label{eqn2.5}
\eeq
The null wave data on ${\cal I}^-$ are: the asymptotic dilatonic wave form $f 
(v,\theta , \varphi)$, and the two polarization components $f_+ (v,\theta , 
\varphi)$, $f_{\times} (v,\theta , \varphi)$, of the asymptotic gravitational 
wave form $f_{\mu \nu} (v,\theta , \varphi)$ \footnote{The three functions $f$, 
$f_+$, $f_{\times}$ of $v,\theta , \varphi$ are equivalent to six functions in 
$\Rb^3$, i.e. six functions of $r,\theta , \varphi$ with $r \geq 0$, because 
the advanced time $v$ ranges over the full line $(-\infty , +\infty)$.}. 
[Introducing a local orthonormal frame ${\bf e}_{(1)}$, ${\bf e}_{(2)}$ on the 
sphere at infinity one usually defines $f_+ = \frac{1}{2} (f_{(1)(1)} - 
f_{(2)(2)})$, $f_{\times} = f_{(1)(2)}$.] The other ${o} (1/r)$ pieces in 
the metric are gauge dependent, except for the ``Bondi mass aspect''  $m_{\infty} 
(v,\theta , \varphi)$ (defined below in a simple case) whose advanced-time 
derivative is related to the (direction-dependent) incoming energy fluxes:
\beq
\frac{\partial}{\partial v} \, m_{\infty} (v,\theta , \varphi) = \frac{1}{4} \, 
(\partial_v \, f)^2 + \frac{1}{4} \, (\partial_v \, f_+)^2 + \frac{1}{4} \, 
(\partial_v \, f_{\times})^2 \, + {\rm div.}, 
\label{eqn2.7}
\eeq
where ${\rm div.}$ denotes an angular divergence, $(\sin \theta)^{-1}\,
\pa_\theta(\sin \theta D^\theta) + \pa_\vphi D^\vphi$.
Instead of the asymptotic wave forms $f$, $f_+$, $f_{\times}$ (which have the 
dimension of length), one can work with the corresponding ``news functions''
\beq
N (v,\theta , \varphi) \equiv \partial_v \, f (v,\theta , \varphi) \ , \ N_+ 
\equiv \partial_v \, f_+ \ , \ N_{\times} \equiv \partial_v \, f_{\times} \, , 
\label{eqn2.8}
\eeq
which are dimensionless, and whose squares give directly the incoming energy 
flux appearing on the R.H.S. of Eq.~(\ref{eqn2.7}).

Before specializing this generic ingoing state, let us note \cite{V97}, 
 the existence of 
two important global symmetries of the classical field equations, 
(\ref{eq1.4}) and (\ref{eqn2.3}). They are invariant both 
under global scale transformations, and under a constant shift of $\phi : ds'^2 
= a^2 ds^2$, $\phi' = \phi + b$. [They are also invariant under local 
coordinate transformations.] In terms of asymptotic data (whose definition 
requires a specific ``flat'' normalization of the coordinates at past 
infinity), the global symmetry transformations read:
\bea
\label{p1}
&&f(v,\theta,\varphi) \rightarrow f'(v',\theta,\varphi) 
= a \, f (v'/a 
,\theta,\varphi) \,,\\
&& f_+(v,\theta,\varphi) \rightarrow f'_+ (v',\theta,\varphi) =
 a \, f_+ (v'/a ,\theta,\varphi)\,, \\
&&f_{\times}(v,\theta,\varphi) \rightarrow f'_{\times} (v',\theta,\varphi) = 
a \, f_{\times} (v'/a,\theta,\varphi)
\label{eqn2.9}
\eea
with $r \rightarrow r^\prime = a r$, $v \rightarrow v^\prime = a v$, and 
\beq
\phi_0 \rightarrow \phi'_0 = \phi_0 + b \, .
\label{eqn2.10}
\eeq
Note that the three dimensionless news functions (\ref{eqn2.8}) are numerically 
invariant under the scaling transformations: $N' (v',\theta,\varphi) = N 
(v,\theta,\varphi)$, etc. with $v' = a \, v$.

As the full time evolution  of string vacua depends only on the null data, 
it is a priori clear that the amplitude of the dimensionless news functions 
must play a crucial r\^ole in distinguishing ``weak" incoming fields --that finally 
disperse again as weak outgoing waves-- from ``strong" incoming fields that undergo 
a gravitational instability. To study this difficult problem in more detail we 
turn in the next section  to a simpler model with fewer degrees of freedom.

Before doing so, let us comment on the ``classicality'' of an 
``in state'' defined by some news functions (\ref{eqn2.8}). The classical 
fields (\ref{eqn2.4}), (\ref{eqn2.5}) deviate less and less from 
a trivial background $(\phi_0, \eta_{\mu \nu})$ as $r \rightarrow 
\infty$ (with fixed $v, \theta$ and $\varphi$). One might then worry 
that, sufficiently near ${\cal I}^-$, quantum fluctuations 
$\phi^q$, $h_{\mu \nu}^q$ might start to dominate over 
the classical deviations $\phi^c \equiv \phi(x)-\phi_0$, 
$h_{\mu \nu}^c \equiv g_{\mu \nu} - \eta_{\mu \nu}$. 
To ease this worry let us compute the mean number of quanta 
in the incoming waves (\ref{eqn2.4}), (\ref{eqn2.5}). Consider 
for simplicity the case of an incoming spherically symmetric dilatonic wave 
$\phi^c_{\rm in} = f(v)/r + o(1/r)$. Such an incoming wave 
has evidently the same mean number of quanta as the corresponding 
outgoing \footnote{In keeping with usual physical intuition, it is 
more convenient  to work with an auxiliary problem where a classical source 
generates an outgoing wave which superposes onto the quantum 
fluctuations of an incoming vacuum state.} 
wave $\phi^c_{\rm out}$, with asymptotic 
behaviour $\phi_{\rm out}^c= -f(u)/r + o(1/r)$ , on ${\cal I}^+$ 
($r \rightarrow \infty$, with $u \equiv t-r, \theta$ and $\varphi$ 
fixed). As is well known \cite{IZ}, a classical outgoing 
wave can be viewed, quantum mechanically, as a coherent 
state in the incoming Fock space defined by the oscillator 
decomposition of the quantum field $\widehat{\phi}$. Instead 
of parametrizing $\phi^c_{\rm out}$ by the waveform $f(u)$, 
one can parametrize it by an effective source $J(x^\lambda)$ 
such that 
$\phi^c_{\rm out}(x) = \int d^4 x^\prime \, P_{\rm ret}(x - x^\prime)\,
J(x^\prime)$, where $P_{\rm ret}$ is the (action-normalized) 
retarded propagator of the field $\phi$. [We use a 
normalization such that $S = -\frac{1}{2}\,\phi\,P^{-1}\,\phi + J\,\phi$.]
The mean number of quanta in the coherent 
state associated to a classical source $J(x)$ 
has been computed in \cite{IZ} for the electromagnetic case, and 
in \cite{BD98} for generic massless fields. It reads 
(with $p\, x \equiv \vec{p}\,\vec{x} - \omega\,t$) 
\vskip 0.3truecm
\beq
\overline{n} = \frac{\pi}{\hbar}\,\int \frac{d^4 p}{(2 \pi)^4}\,
\delta(p^2)\,\widehat{J}(-p)\,{\cal R}\,\widehat{J}(p)\,, \quad \quad 
\widehat{J}(p) \equiv \int d^4 x \, e^{-i p x}\, J(x)\,, 
\label{n1}
\eeq
\vskip 0.3truecm
\noindent
where ${\cal R}$ denotes the residue of the propagator of the corresponding 
field, defined such that the field equation $P^{-1}\,\phi = J$ reads 
$\Box \phi = - {\cal R}\,J$. [In the normalization of the present paper 
${\cal R}=16 \pi\,G$.] Inserting $J(x) = -(4\,G)^{-1}\,
f(t)\,\delta^{(3)}(\vec{x})$ into Eq.~(\ref{n1}) 
yields 
\vskip 0.3truecm
\beq
\overline{n} = \frac{1}{2 G\hbar} \, \int_0^\omega \frac{d \omega}{2 \pi} 
\, \omega\,|\widehat{f}(\omega)|^2 = 
\frac{1}{2 G\hbar} \, \int_0^\omega \frac{d \omega}{2 \pi} 
\, \frac{1}{\omega}\, |\widehat{N}(\omega)|^2\,,
\eeq
\vskip 0.3truecm
\noindent
where $\widehat{N}(\omega) \equiv \int dt \, e^{i \omega t}\, N(t)$ 
is the one-dimensional Fourier transform of the news function 
$N(t) = f^\prime(t)$. In order of magnitude, if we consider 
a pulse-like waveform $f(t)$, with characteristic duration $\Delta t 
\sim \ell_c$, and characteristic dimensionless news amplitude 
$N_c \sim f_c/\ell_c$, the corresponding mean number of quanta is 
\beq
\overline{n} \sim \frac{N_c^2\,\ell_c^2}{\ell_P^2} = 
\frac{N_c^2\,\ell_c^2}{g^2\,\ell_s^2}\,.
\label{n3}
\eeq
In the present paper, we shall consider incoming news functions with 
amplitude $N_c \sim 1$ and scale of variation $\ell_c \gg \ell_s$. 
Moreover, in order to ensure a sufficient amount of inflation, 
the initial value of the string coupling $g^2 = e^{\phi_0}$ 
has to be $\ll 1$. Therefore, Eq.~(\ref{n3}) shows that such incoming states 
can be viewed as highly classical coherent states,  
made of an enormous number  of elementary $\phi$-quanta: 
$\overline{n} \gg\!\!> 1$. This quantitative fact clearly shows 
that there is no need to worry about the effect of quantum fluctuations, 
even near ${\cal I}^-$ where $\phi^c_{\rm in } \sim f(v)/r \rightarrow 0$. 
In the case of the negative-curvature Friedmann-dilaton solution 
this fact has been recently confirmed by an explicit calculation 
of scalar and tensor quantum fluctuations \cite{GPV}.

\section{Equations for a spherically symmetric pre-big bang}
\label{sec3}
The simplest nontrivial model which can exhibit the transition between weak, 
stable data, and gravitationally unstable ones is the spherically-symmetric 
Einstein-dilaton system. The null wave data for this system comprise only an 
angle-independent asymptotic dilatonic wave form $f(v)$, corresponding to the 
dimensionless news function
\beq
N(v) = \partial_v \, f(v) \, . 
\label{eqn3.1}
\eeq
The spherically symmetric Einstein-dilaton system has been thoroughly studied 
by many authors with the strongest analytic results appearing 
in papers by Christodoulou \cite{chi86}--\cite{chi93}. 
A convenient system of coordinates is the double null 
system, $(u,v)$, such that
\beq
\phi = \phi (u,v) \, , 
\label{eqn3.2}
\eeq
\beq
ds^2 = - \Omega^2(u,v)\,du \, dv + r^2(u,v)\,d \omega^2 \,,
\label{eq1.6}
\eeq
where $d\omega^2 = d\theta^2 + \sin^2 \theta \, d\varphi^2$. The field 
equations are conveniently re-expressed in terms of the three functions $\phi 
(u,v)$, $r(u,v)$ and $m(u,v)$, where the local mass function $m(u,v)$ is 
defined by \footnote{
In order to avoid confusion we indicate, in standard notation, the variable 
kept fixed under differentiation.} 
\beq
1 - \frac{2 m}{r} \equiv g^{\mu \nu}\,(\pa_\mu r)\,(\pa_\nu r) = 
-\frac{4}{\Omega^2}\,\left (\frac{\pa r}{\pa u}\right )_v\,
\left (\frac{\pa r}{\pa v}\right )_u\,.
\label{eq1.7}
\eeq
One gets the following  set of evolution equations for $m$, $r$ and $\phi$ 
\bea
\label{eq1.8}
&& 2\left (\frac{\pa r}{\pa u}\right )_v
\,\left (\frac{\pa m}{\pa u}\right )_v = \left ( 1 - \frac{2 m}{r} \right )
\,\frac{r^2}{4}\,\left (\frac{\partial \phi}{\pa u} \right )_v^2 \,, \\
\label{eq1.9}
&& 2 \left (\frac{\pa r}{\pa v}\right )_u\,
\left (\frac{\pa m}{\pa v}\right )_u = \left ( 1 - \frac{2 m}{r} \right )
\,\frac{r^2}{4}\,\left (\frac{\partial \phi}{\pa v} \right )^2_u \,,\\
\label{eq1.10}
&& r\,\frac{\pa^2 r}{\pa u \pa v} = 
\frac{2 m}{r-2m}\,\left (\frac{\pa r}{\pa u}\right )_v\,
\left (\frac{\pa r}{\pa v}\right )_u\,,\\
\label{eq1.11}
&& r\,\frac{\pa^2 \phi}{\pa u \pa v} + 
 \left (\frac{\pa r}{\pa u}\right )_v \left (\frac{\pa \phi}{\pa v}\right )_u
+ \left (\frac{\pa r}{\pa v}\right )_u \left (\frac{\pa \phi}{\pa u}\right )_v 
=0\,.
\eea
This double-null evolution system is form-invariant under independent local 
reparametrizations of $u$ and $v$: $u' = U(u)$, $v' = V(v)$. To freeze this 
gauge freedom it is convenient to require that $u=v$ on the central worldline 
of symmetry $r=0$ (as long as it is a regular part of space-time), and that $v$ 
and $r$ asymptotically behave like ingoing Bondi coordinates on ${\cal I}^-$ 
(see below).

The quantity
\beq
\mu (u,v) \equiv \frac{2 m (u,v)}{r}
\label{eqn3.3}
\eeq
plays a crucial r\^ole in the problem. 
Following \cite{chi93}, we shall sometimes call it the ``mass 
ratio''. If $\mu$ stays everywhere below 1, the field configuration will 
not collapse but will finally disperse itself as outgoing waves at infinity. By 
contrast, if the mass ratio $\mu$ can reach anywhere the value 1, this signals 
the formation of an apparent horizon ${\cal A}$. The location of the apparent 
horizon is indeed defined by the equation
\beq
{\cal A}: \quad \mu (u,v) = 1 \, .
\label{eqn3.4}
\eeq
The above statements are substantiated by some rigorous inequalities \cite{chi93}
stating that:
\bea
\label{rel1}
&& \left (\frac{\pa r}{\pa u}\right )_v < 0\,, \quad \quad \quad \quad 
\quad \left (\frac{\pa m}{\pa v}\right )_u \geq 0\,, \\
\label{rel2}
&& \left (\frac{\pa r}{\pa v}\right )_u\,
\left ( 1- \mu \right ) \geq 0\,, \quad \quad 
\left (\frac{\pa m}{\pa u}\right )_v\,
\left ( 1- \mu \right ) \leq 0\,.
\eea
Thus, in weak-field regions ($\mu < 1$), $(\partial_v 
r)_u > 0$, while, as $\mu > 1$, $(\partial_v r)_u < 0$, meaning that the 
outgoing radial null rays (``photons'') emitted by the sphere $r = {\rm const.}$ 
become convergent, instead of having their usual behaviour. 

\begin{figure}
\centerline{\epsfig{file=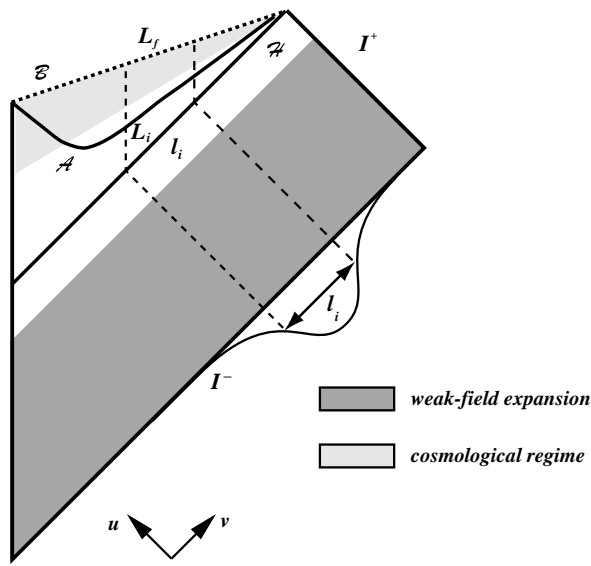,width=0.45\textwidth,angle=-90}}
\vskip 0.3truecm
\caption{\sl Schematic representation of the space-time generated 
by the collapse of a spherically symmetric pulse of dilatonic 
waves. An incoming pulse of scalar news $N(v)$, which grows 
by ${\cal O}(1)$ on an advanced-time scale $\ell_i$, collapses 
to a space-like singularity ${\cal B}$ after having formed 
an apparent horizon ${\cal A}$ hidden behind the event horizon ${\cal H}$. 
In the vicinity of ${\cal A}$ and ${\cal H}$, there is an abrupt 
transition between a weak-field region (dark shading) where the 
perturbation series (\ref{eqn4.1a})--(\ref{eqn4.1c}) holds, and 
a strong-field one (light shading) where the cosmological-like 
expansion (\ref{appen1})--(\ref{appen3}) holds.}
\label{fig3}
\end{figure}

It has been shown long ago that the presence of trapped surfaces 
implies the existence of some (possibly weak) type of 
geometric singularity \cite{P65}, \cite{HP70}. In the case of the spherically 
symmetric Einstein-dilaton system, it has been possible to prove 
\cite{chi91} that the 
presence of trapped surfaces (i.e. of an apparent horizon ${\cal A}$ (\ref{eqn3.4}), 
the boundary of the trapped region) implies the existence of a future 
space-like singular boundary ${\cal B}$ of space-time where the curvature blows up.
Both ${\cal A}$ and ${\cal B}$ are ``invisible'' from future null infinity 
(${\cal I}^+$), being hidden behind an event horizon 
${\cal H}$ (a null hypersurface). See Fig.~\ref{fig3}.

One of our main purposes in this work is to give the conditions that the incoming 
dilatonic news function $N(v)$ must satisfy in order to create an apparent horizon, and 
thereby to lead to some localized gravitational collapse. Before addressing this 
problem let us complete the description of the toy model by re-expressing it
 in terms of the ingoing Bondi coordinates $(v,r)$. The 
metric can now be written in the form 
\beq
\label{eq1.17}
ds^2 = - \left (1 - \frac{2 m(v,r)}{r} \right )\,e^{2 \beta(v,r)}\,d v^2 +
2 e^{\beta(v,r)}\, dv \, dr + r^2 \, d \omega^2\,,
\eeq
and the field equations (\ref{eq1.8})--(\ref{eq1.11}) become 
\bea
\label{eq1.18}
&& r\,\left (\frac{\pa \beta}{\pa r}\right )_v  = \frac{r^2}{4}\,
\left (\frac{\pa \phi}{\pa r}\right )_v^2 \,,\\
\label{eq1.19}
&& 2\left (\frac{\pa m}{\pa r}\right )_v = \left ( 1 - \frac{2 m}{r} \right )\,
\frac{r^2}{4}\,\left (\frac{\partial \phi}{\pa r} \right )^2_v \,,
\eea
\beq
\label{new}
2 \left (\frac{\pa m}{\pa v}\right )_r = 
e^{-\beta}\,\frac{r^2}{2}\,\left [ \left (\frac{\pa \phi}{\pa v}\right )_r^2 + 
e^\beta\,\left ( 1 - \frac{2 m}{r} \right )\, 
\left (\frac{\pa \phi}{\pa r}\right )_v\,
\left (\frac{\pa \phi}{\pa v}\right )_r \right ]\,,
\eeq
\beq
\label{eq1.20}
2 \left [ \frac{\pa^2 \phi}{\pa v \pa r } + \frac{1}{r}\,
 \left (\frac{\pa \phi}{\pa v}\right )_r \right ] + e^\beta \left (1-\frac{2m}{r} \right )\,
 \left ( \frac{\pa^2 \phi}{\pa r^2}\right )_v  +{ 2e^\beta \over r} \,
\left (1-\frac{m}{r} \right)
 \left (\frac{\pa \phi}{\pa r} \right )_v  = 0\,.
\eeq
Eqs. (\ref{eq1.18}), (\ref{eq1.19}) imply: 
\beq
\pa_r \left (e^{\beta(v,r)}\, ( r - 2 m ) \right )= e^{\beta(v,r)}\,.
\eeq 
In this coordinate system one can solve for the two functions $\beta (v,r)$ and 
$\mu (v,r)$ (or $m (v,r)$) by quadratures in terms of the third unknown 
function $\phi (v,r)$. Indeed, imposing that the coordinate 
system be asymptotically flat at ${\cal I}^-$ (i.e. that $\beta (v,r) 
\rightarrow 0$ as $ r \rightarrow \infty$), one first finds from 
Eq.~(\ref{eq1.18}) 
\beq
\beta (v,r) = - \int_r^{+\infty} \frac{dr'}{r'} \, F (v,r') \, ,
\label{eqn3.5}
\eeq
where 
\beq
\label{eq1.24}
F(v,r) \equiv \frac{r^2}{4}\,\left ( \frac{\pa \phi}{\pa r} \right )_v^2\,.
\eeq
Eq.~(\ref{eq1.19}) can then be integrated to give 
\beq
\mu(v,r) = \frac{\left [ 2m_{\infty}(v) - 
\int_r^{+\infty} dr^\prime \, F(v,r^\prime)\,e^{\beta(v,r^\prime)} 
\right ]}{r \,e^{\beta(v,r)}}\,,
\label{eqn3.6}
\eeq
where the ``integration constant''
\bea
m_{\infty} (v) \equiv \lim_{r \rightarrow \infty} m (v,r) 
\eea
denotes the incoming Bondi mass. By the Bondi energy-flux formula 
(\ref{eqn2.7}), which is also the limiting form of (\ref{new}) for 
$r \rightarrow \infty$, 
$m_{\infty} (v)$ is given in terms of the news function by
\beq
m_{\infty} (v) = \frac{1}{4} \int_{-\infty}^v dv' \, N^2 (v') =  \, \frac{1}{4} 
\int_{-\infty}^v dv' \left( \frac{\partial f (v')}{\partial v'} \right)^2 \,,
\label{eqn3.7}
\eeq
where we have inserted the asymptotic behaviour of $\phi$ on ${\cal I}^-$,
\beq
\phi (v,r) = \phi_0 + \frac{f(v)}{r} + 
{o} \left( \frac{1}{r} \right) \,,
\label{eqn3.8}
\eeq
and assumed  that $N(v)$ decays sufficiently fast as $v \rightarrow 
-\infty$ (i.e. near past time-like infinity). 
{}From Eqs.~(\ref{eqn3.5}) and (\ref{eqn3.6}) we find the leading asymptotic 
behaviour of the metric coefficients to be
\beq
1 - \frac{2m (v,r)}{r} = 1 - \frac{2 m_{\infty} (v)}{r} +  {o} \left( 
\frac{1}{r} \right) \, ,
\label{eqn3.9a}
\eeq
\beq
e^{\beta (v,r)} = 1 + {\cal O} \left( \frac{1}{r^2} \right) \, .
\label{eqn3.10}
\eeq

\section{Data Strength criterion for gravitational instability}
\label{sec4}
The purpose of this section is to outline the condition that the initial 
data, i.e. the wave form $f(v)$ or the news function $N(v) \equiv f' (v)$, must 
satisfy in order to undergo, or not undergo, gravitational collapse. A lot of 
mathematical work has been done on this issue \cite{chi86}--\cite{chi93}. 
In particular, Refs.~\cite{chi86} and \cite{chi93} 
gave two different ``no collapse'' criteria 
ensuring that the field 
configuration never collapses and finally disperses out as weak outgoing waves
if some functional of the data is small enough. 
On the other hand, Ref.~\cite{chi91} gave a sufficient ``collapse'' criterion 
ensuring that the field 
configuration will collapse, if some functional of the data is large enough,
thereby giving birth to a curvature singularity 
hidden behind a horizon. The problem with these nice and important 
mathematical results is threefold: (i) these criteria are 
sufficient but not necessary, so that they cannot answer our problem of finding 
(if possible) a sharp criterion distinguishing weak-field from strong-field data; 
(ii) the various measures of the ``strength'' of the data given in 
Refs.~\cite{chi86}, \cite{chi91} and \cite{chi93} are quite unrelated to each 
other and do not point clearly toward any sharp ``strength criterion''; and 
(iii) these criteria are not expressed in terms of the asymptotic
null data $N(v)$.

Our aim here is not to compete with Refs.~\cite{chi86}, \cite{chi91}, 
\cite{chi93} on the grounds of mathematical rigour, but to complement these 
results by a non rigorous study which leads to the iterative computation of a 
sharp strength criterion, i.e. a functional ${\cal S} [f]$, such that data 
satisfying ${\cal S} < 1$ finally disperse out at infinity, while data 
satisfying ${\cal S} > 1$ are gravitationally unstable and partly collapse to 
form a singularity. We will then compare our strength functional ${\cal S}$ 
with the rigorous, but less explicit results of \cite{chi91}. Note that the 
emphasis here is not on the question of whether the created singularity is hidden 
beyond a global event horizon when seen from outside, but rather on the 
quantitative criterion ensuring that past trivial data give birth to a 
space-like singularity which we shall later interpret as a budding pre-Big Bang 
cosmological universe.

\subsection{ Perturbative analysis in the weak-field region}
\label{subsec3.1}

In view of the discussion of the previous section, the transition between 
``weak" and ``strong" fields is sharply signaled by the appearance of an 
apparent horizon, i.e. of space-time points where $\mu \equiv 2m/r$ reaches the 
``critical" value 1. Therefore, starting from the weak-field region near ${\cal 
I}^-$, we can define the strength functional by computing $\mu (u,v)$ as a functional 
of the null incoming data and by studying if and when it can exceed the critical value. 
Specifically, we 
can set up a perturbation analysis of the Einstein-dilaton system in some 
weak-field domain connected to ${\cal I}^-$, and see for what data it suggests 
that $\mu$ will exceed somewhere the value 1. The expected domain of validity 
of such a weak-field expansion is sketched in Fig.~\ref{fig3}.

The $(u,v)$ system, Eqs.~(\ref{eqn3.2}) and (\ref{eq1.6}),
lends itself easily to a perturbative treatment in the 
strength of $\phi$. We think here of introducing a formal parameter, say 
$\lambda$, in the asymptotic data by $f(v) \rightarrow \lambda \, f(v)$ and of 
constructing the full solution $\phi (u,v)$, $r(u,v)$, $m(u,v)$, as a (formal) 
power series in $\lambda$ (or by some better iteration method):
\bea
\label{eqn4.1a}
\overline{\phi} (u,v) &\equiv& \phi (u,v) - \phi_0 = \lambda \, \phi_1 + 
\lambda^3 \phi_3 + \lambda^5 \phi_5 + \ldots \,, \\
\label{eqn4.1b}
r(u,v) &=& \frac{1}{2} \, (v-u) + \lambda^2 \, r_2 + \lambda^4 \, r_4 + \ldots 
\,, \\
m(u,v) &=& \lambda^2 \, m_2 + \lambda^4 \, m_4 + \ldots \, .
\label{eqn4.1c}
\eea
Indeed, knowing $\phi$ to order $\lambda^{2p-1}$ and $r$ to order 
$\lambda^{2p-2}$ we can use Eq.~(\ref{eq1.8}) or (\ref{eq1.9}) to compute by 
quadrature $m$ to order $\lambda^{2p}$. We can then  rewrite Eqs.~(\ref{eq1.10}), 
(\ref{eq1.11}) as
\beq
\partial_{uv} (r \, \overline{\phi}) = \omega \,r \, \overline{\phi}\,, 
\quad \quad \partial_{uv} \, r =\omega \, r \,, 
\label{eqn4.2}
\eeq
where
\beq
\omega \equiv \frac{\mu}{1-\mu} \, \frac{\partial_u r \, \partial_v r}{r^2} \, , 
\label{eqn4.3}
\eeq
and therefore the right-hand sides of Eqs.~(\ref{eqn4.2}) are known to 
(relative) order $\lambda^{2p}$. Solving Eq.~(\ref{eqn4.2}), then yields 
$\overline{\phi}$ and $r$ to the next order in $\lambda$ ($\lambda^{2p+1}$ and 
$\lambda^{2p}$, respectively). To be more explicit, Eqs.~(\ref{eqn4.2}) 
are of the form
\beq
\partial_{uv} \, \psi(u,v) = \sigma (u,v) \, , 
\label{eqn4.4}
\eeq
where the ``source'' $\sigma = \omega \psi$ is known at each order in $\lambda$ 
from the lower-order expression of $\mu$ and $\psi$, and where the ``field'' 
$\psi (u,v)$ is restricted to vanish on the central worldline of symmetry $v=u$ 
(where $r=0$, and $\overline{\phi}$ is regular in the weak-field domain). The 
generic Eq.~(\ref{eqn4.4}) can then be solved by introducing the retarded 
Green function
\bea
G (u,v ; u',v') = \theta (u-u') \, \theta (v-v') - \theta (v-u') \, \theta 
(u-v')  \, ,
\eea
\beq
\partial_{uv} \, G (u,v ; u',v') = \delta (u-u') \, \delta (v-v') \, . 
\label{eqn4.5}
\eeq
This  is the unique Green function in the domain $v \geq u$, $v' \geq u'$ which 
vanishes at $v = u$, and whose support in the source point 
$(u',v')$ lies in the past of the field point $(u,v)$. The general solution of 
(\ref{eqn4.4}) can then be written as
\beq
\psi (u,v) = \psi_{\rm in} (u,v) + \int \int_{v' \geq u'} du' \, dv' \, G (u,v 
; u',v') \, \sigma (u',v') \, ,
\label{eqn4.6}
\eeq
where $\psi_{\rm in} (u,v)$ is the free ``incoming'' field, satisfying 
$\partial_{uv} \, \psi_{\rm in} = 0$ and vanishing when $v=u$, i.e.
\beq
\psi_{\rm in} (u,v) = g_{\rm in} (v) - g_{\rm in} (u) \, .
\label{eqn4.7}
\eeq
The incoming waveform $g_{\rm in} (v)$ is uniquely fixed by the incoming data 
on ${\cal I}^-$. For instance, for $r(u,v)$ asymptotic flatness gives $r_{\rm 
in} (u,v) = \frac{1}{2} \, (v-u)$ (we recall that in flat space $v = t+r$ and 
$u=t-r$), while for $\overline{\phi} (u,v) \equiv \phi (u,v) - \phi_0$ the 
asymptotic expansion (\ref{eqn3.8}) (in which $f(v)$ was supposed to vanish as 
$v \rightarrow -\infty$) yields $\overline{\phi}_{\rm in} (u,v) = 2(f(v) - f(u))/(v-u)$.

In principle, this perturbative algorithm allows one to compute the mass ratio 
$\mu (u,v) = 2m/r$ to any order in $\lambda$:
\beq
\mu (u,v) = \lambda^2 \, \mu_2 + \lambda^4 \, \mu_4 + \ldots\,.
\label{eqn4.8}
\eeq
Evidently, the convergence of this series becomes very doubtful when $\mu$ can 
reach values of order unity, which is precisely when an apparent horizon is 
formed. One would need some resummation technique to better locate the apparent 
horizon condition $\mu (v,u) = 1$. However, we find interesting to have, in 
principle, a way of explicitly computing successive approximations to 
the possible location of an apparent horizon, starting only from the incoming 
wave data $f(v)$. 

\subsection{Strength criterion at quadratic order} 
\label{subsec3.2}
Let us compute explicitly the lowest-order approximation to 
(\ref{eqn4.8}) (we henceforth set 
$\lambda = 1$ for simplicity). It is obtained by inserting the zeroth-order result $r_0 (u,v) 
= \frac{1}{2} \, (v-u)$, and the first-order one for $\phi$
\beq
\overline{\phi}_1 (u,v) = \phi_1 (u,v) - \phi_0 = \frac{f(v) - f(u)}{r_0} =  
\, \frac{2[f(v) - f(u)]}{ v-u}
\label{eqn4.9}
\eeq
into Eq.~(\ref{eq1.8}) or (\ref{eq1.9}). This yields (with $f' (v) \equiv 
\partial f / \partial v$)
\bea
\left( \frac{\partial m_2}{\partial v} \right)_u = r_0^2 \left( \frac{1}{2} \, 
\frac{\partial \phi_1}{\partial v} \right)^2 = \frac{1}{4} \left [ f'(v) - 
\frac{f(v) - f(u)}{v-u} \right ]^2 \,,
\eea
\beq
\left( \frac{\partial m_2}{\partial u} \right)_v = - r_0^2 \left( \frac{1}{2} 
\, \frac{\partial \phi_1}{\partial u} \right)^2 = -\frac{1}{4} \left [ f'(u) -
\frac{f(v) - f(u)}{v-u} \right ]^2 \, .
\label{eqn4.10}
\eeq
The compatibility of the two equations is easily checked, e.g. by using
\beq
\partial_v (r_0^2 (\partial_u \phi_1 )^2) = r_0 \, \partial_u \phi_1 \, 
\partial_v \phi_1 = - \partial_u (r_0^2 (\partial_v \phi_1)^2) \, ,
\label{eqn4.11}
\eeq
which expresses the ``conservation'' of the $\phi$-energy tensor in the $(u,v)$ 
plane. Noting that $m(u,v)$ must vanish (by regularity of $r=0$ in the 
weak-field domain) at the center $v=u$ one gets, by quadrature, the following 
explicit result for $m_2$:
\beq
m_2 (u,v) = \widetilde m (u,v) - \frac{1}{8} \ r_0 \, \overline{\phi}_1^2 \, ,
\label{eqn4.12}
\eeq
where 
\beq
\widetilde m (u,v) \equiv \frac{1}{4} \int_u^v f'^2 (x) \, dx \, .
\label{eqn4.13}
\eeq
Note that Eq.~(\ref{eqn4.12}) decomposes $m_2 (u,v)$ in a ``conserved'' piece 
$\widetilde m (u,v) = g(v) - g(u)$ (with $\partial_{uv} \widetilde m = 0$) and 
a $\phi$-dependent one. This is an integrated form of the local 
``conservation'' law (\ref{eqn4.11}) of the $\phi$-energy tensor. Using 
(\ref{eqn4.12}) one gets a very simple result for the quadratic approximation 
to the mass ratio $\mu = 2m/r$, namely
\beq
\mu_2 (u,v) = \frac{1}{v-u} \int_u^v dx \, f'^2 (x) - \left [ 
\frac{f(v) - f(u)}{v-u} \right ]^2 \, .
\label{eqn4.14}
\eeq
Let us introduce a simple notation for the {\it average} of any function $g(x)$ 
over the interval $[u,v]$:
\beq
\langle g \rangle_{[u,v]} \equiv \langle g (x) \rangle_{x \in [u,v]} \equiv 
\frac{1}{v-u} \int_u^v dx \, g(x) \, .
\label{eq3.8}
\eeq
Then the mass ratio, at quadratic order, can be simply written in terms of the 
scalar news function $N(v) \equiv f'(v)$:
\bea
\mu_2 (u,v) &= &\langle N^2 \rangle_{[u,v]} - (\langle N \rangle_{[u,v]})^2\,, \\
&= &\langle (N(x) - \langle N \rangle_{[u,v]})^2 \rangle_{x \in [u,v]}\,, \\
&\equiv &{\rm Var} \, (N(x))_{x \in [u,v]} \, ,
\label{eqn4.15}
\eea
where ${\rm Var} \, (g)_{[u,v]}$ denotes the ``variance'' of the function 
$g(x)$ over the interval $[u,v]$, i.e. the average squared deviation from the 
mean.

As indicated above, having obtained $\mu$ to second order, say $\mu_2 = V(u,v) 
\equiv {\rm Var} \, (N)_{[u,v]}$, one can then proceed to compute $r$ and 
$\phi$ to higher orders. Namely, from Eq.~(\ref{eqn4.6})
\beq
2r(u,v) = v-u - G * \left [ \frac{V(u,v)}{v-u} \right ] + {\cal O} (\lambda^4) \,, 
\eeq
\beq
(r\phi)(u,v) = f(v) - f(u) - G * \left [ V(u,v) \, \frac{f(v) - f(u)}{(v-u)^2} 
\right ] + {\cal O} (\lambda^5) \,, 
\eeq
where the star denotes a convolution. Then, one can get from these results 
$m_4$, and thereby $\mu_4$, by simple quadratures. In the following we shall 
only use the quadratic approximation to $\mu$, though we are aware that when 
the initial data are strong enough to become gravitationally unstable, higher 
order contributions to $\mu$ are probably comparable to $\mu_2$.

Finally, at quadratic order, we can define the strength of some initial data 
simply as the supremum of the variance of the news over arbitrary intervals:
\beq
{\cal S}_2 = \sup \mu_2 = \sup_{u,v \atop u \leq v} \, {\rm Var} (N(x))_{x \in [u,v]} \, .
\label{eqn4.16}
\eeq
At this order ${\cal S}_2 < 1$ means that $\mu_2 (u,v)$ stays always below one, 
i.e. that no trapped region is created, and that the field is expected to 
disperse, while ${\cal S}_2 > 1$ signals the formation, somewhere, of trapped 
spheres, and therefore (by the results of \cite{chi91}) the formation of a 
singularity. To be more exact we should actually define,  in analogy with
(\ref{eqn4.16}),
 a quantity ${\cal S}_{2p} = \sup \sum_{k=1}^p \mu_{2k}$, 
at any finite order in
the weak field expansion, and state  our collapse criterion as the inequality:
\beq
{\cal S}_{2p} > C_{2p}\,, \quad \quad C_{2p} \stackrel{p\rightarrow \infty}{\rightarrow} 1\,.
\label{gencrit}
\eeq
In practice, even if
 we do not expect it to be 
quantitatively exact, we will use  the criterion (\ref{gencrit}) with $p=1$, taking
$C_2 =1$,  hoping that it will be qualitatively correct in 
capturing the features of the news function which are generically 
important for producing a gravitational collapse. Let us note that the 
functional ${\cal S}_2$ is (like $N$) dimensionless (and therefore scale 
invariant) and that it is nonlocal. We note also that it is invariant under a 
constant shift of $N$, $N(v) \rightarrow N(v) + a$, which corresponds to adding 
a linear drift in $f : f(v) \rightarrow f(v) + av + b$. Such a shift is 
(formally) equivalent, in view of Eq.~(\ref{eqn4.9}), to a constant shift of 
$\phi_0 \rightarrow \phi_0 + 2a$.

The non locality of ${\cal S}_2$ is physically interesting because it indicates 
that it is not the instantaneous level of the energy flux $N^2 (v)$ which 
really matters, but rather the possibility of having a flux which varies by $\sim$ 
100\% over some interval of advanced time. This non locality defines also some 
characteristic scales associated with the collapse (when ${\cal S}_2 > 1$). 
Indeed, if, by causality, we consider increasing values of $v$, and define 
${\cal S}_2 (v) = {\displaystyle \sup_{u,u \leq v}} {\rm Var} (N)_{[u,v]}$, the 
first value of $v$ for which ${\cal S}_2 (v)$ exceeds $1$ defines an advanced 
characteristic time when the collapse occurs, and the corresponding maximizing 
interval $[u,v]$ defines a characteristic time scale of (in)homogeneity. [Note 
that $\mu_2 (u,v)$, Eq.~(\ref{eqn4.15}), vanishes when $u=v$, and vanishes also 
generally when $u \rightarrow -\infty$ if $N(u)$ tends to a limit at 
$-\infty$.]

Let us also mention that our strength functional (\ref{eqn4.16}) is 
superficially similar to the one recently introduced by Christodoulou 
\cite{chi93} to characterize sufficiently weak (i.e. non collapsing) data. 
In the lowest order 
approximation his criterion measures the strength of the data 
by ${\cal V} = TV (N(x))$, where $TV$ denotes the total variation (i.e. 
essentially ${\cal V} = \int_{-\infty}^{+\infty} dv \vert N' (v) \vert$). Like 
our variance, this is a measure of the variation of $N(v)$ with, however, a 
crucial difference. If ${\cal V}$ were a good criterion for measuring the 
strength of possibly strong data, we would conclude that a news function which 
oscillates with a very small amplitude for a very long time will be 
gravitationally unstable, while our criterion indicates that it will not, which 
seems physically more plausible. We hope that our strength functional 
(\ref{eqn4.16}) will suggest new gravitational stability theorems to 
mathematicians.
\subsection{Comparison with a collapse criterion of Christodoulou}
\label{subsec3.3}
To check the reasonableness of our quadratic strength criterion (\ref{eqn4.16}), (\ref{gencrit}) 
we have compared it with the rigorous, but only sufficient, collapse criterion 
of Christodoulou \cite{chi91}, and we have applied it to two simple exact 
solutions. Ref.~\cite{chi91} gives the following sufficient criterion on the strength of 
characteristic data considered at some finite retarded time $u$
\beq
\frac{2 \Delta m}{\Delta r }  \geq 
\left [ \frac{r_1}{r_2}\,\log \left (\frac{r_1}{2 \Delta r} \right )
+ \frac{6r_1}{r_2} -1 \right ]\,,
\label{eq2.2}
\eeq
where $r_1 \leq r_2, r_2 \leq 3r_1/2$, are two spheres, 
$\Delta r = r_2 - r_1$ is the width of the ``annular'' region between the two 
spheres, and $\Delta m = m_2 - m_1 \equiv m(u,r_2) - m (u,r_1)$ is the mass 
``contained'' between the two spheres, i.e. more precisely the energy flux 
through the outgoing null cone $u = {\rm const.}$, between $r_1$ and $r_2$. 
Explicitly, from Eq.~(\ref{eq1.19}) in $(u,r)$ coordinates, we have 
\beq
2 \Delta m = \int_{r_1}^{r_2} dr \, \frac{r^2}{4}\,
\left (1 - \frac{2m}{r} \right )\,
\left (\frac{\pa \phi}{\pa r} \right )^2_{u}\,.
\label{eq2.1}
\eeq
We can approximately express this criterion in terms of the incoming 
null data $N(v) = f'(v)$ if we assume that the outgoing 
cone $u = {\rm const.}$ is in the weak-field domain, so that 
we can replace $\phi$ on the R.H.S. of Eq.~(\ref{eq2.1})
by $\phi = r^{-1} [f(u+2r) - f(u)]$. Then the  
L.H.S. of Eq.~(\ref{eq2.2}) becomes (at quadratic order)
\beq
\frac{2 \Delta m}{\Delta r} = \frac{M(u,v_2) - M(u,v_1)}{v_2-v_1}
\eeq
where 
\beq
\label{mm}
M(u,v) \equiv (v-u)\,{\rm Var}(N)_{[u,v]} 
\equiv \int_u^v d x\, {\left ( N(x) - \langle N \rangle_{[u,v]} \right )}^2\,.
\eeq
Therefore, in this approximation, the criterion of Ref.~\cite{chi91} 
becomes (with $v_0 = u$)
\beq
\exists v_0, v_1, v_2 : \quad \frac{M(v_0,v_2) - M(v_0,v_1)}{v_2-v_1}
\geq C(v_0,v_1,v_2) \,,
\label{chi}
\eeq
with the constraints $2 v_2 - 3v_1 \leq -v_0$, 
$v_1 \leq v_2$, and the definition 
\beq
C(v_0,v_1,v_2) \equiv  \frac{v_1-v_0}{v_2 -v_0}\, 
\log \left (\frac{v_1-v_0}{2(v_2 -v_1)} \right ) + 
\frac{6(v_1-v_0)}{v_2 -v_0} -1\,.
\eeq
If $v_0$ is at the boundary of its allowed domain, 
i.e. if $v_0 = 3v_1 - 2v_2$, this criterion is fully compatible 
with ours since $C(v_0,v_1,v_2)=3$ implying through Eq.~(\ref{mm}) 
\beq
{\var}(N)_{[v_0,v_2]} \geq 1 + \frac{2}{3}\,{\var}(N)_{[v_0,v_1]} \geq 1\,.
\eeq
In the opposite case ($v_0$  large and negative) we think that  the 
criterion (\ref{chi}) is also compatible, for generic news functions,
 with  the general 
form suggested  above, i.e. 
\beq
\exists \, v_1^\prime  , v_2^\prime : \quad  
{\var}(N)_{[v_1^\prime,v_2^\prime]} = \langle N^2 \rangle_{[v_1^\prime,v_2^\prime]}
- \langle N \rangle_{[v_1^\prime,v_2^\prime]}^2 \geq C_2\,,
\label{eqn4.22}
\eeq
with some positive constant $C_2$ of order unity.
We first note that one should probably impose the physically 
reasonable condition that $N(v)$ decays \footnote{
Actually, when comparing our criterion 
(based on asymptotic null data) to that of Ref.~\cite{chi93}
(based on characteristic data taken at a finite retarded 
time $u = v_0$) we can consider, without loss of generality, 
that the news function $N(v)$ vanishes identically for $v \leq v_0$.} 
sufficiently fast as 
$v \rightarrow - \infty$ (faster than $|v|^{-1/2}$) 
to ensure that the integrated incoming 
energy flux $\int_{-\infty}^{v_0} dv \, N^2(v)$ is finite. [This 
constraint freezes the freedom to shift $N(v)$ by a constant.]

When $-v_0$ is large, the question remains, however, to know 
how fast the function  $N(v)$ decays when $v_1-v$ (for $v_0 < v < v_1$)
becomes much larger than $v_2-v_1$. Let us first consider the physically 
generic case where $N(v)$ decays in a reasonably fast manner, say faster 
than a power $|v|^{-\kappa}$ with $\kappa$ of order unity.
Then choosing $v_0$ in the criterion (\ref{chi}) just large enough 
to allow one to neglect $N(v_0)$ with respect to $N(v)$, 
say $(v_1-v_0) \sim (v_2-v_1) \times 10^{1/\kappa}$, 
corresponding to $N(v_0) = 0.1\,N(v_1)$, the L.H.S. 
of Eq.~(\ref{chi}) becomes approximately 
\beq 
\frac{2 \Delta m}{\Delta r} \simeq \langle N^2 \rangle_{[v_1,v_2]}\,,
\label{eqn4.21}
\eeq
while the function $C$ on the R.H.S., which grows only 
logarithmically with $v_0$, becomes  
\beq
C(v_0,v_1,v_2) \simeq \log \left (\frac{v_1-v_0}{2(v_2-v_1)}\right ) +5\,.
\eeq
Therefore, a function of the type $N_\kappa(v) 
\sim c\left [ (v_2-v)/(v_2-v_1)\right]^{-\kappa}$ (when $ v \leq v_1 < v_2$) with 
$c^2 \,\gaq\, 5 + (\log 10)/\kappa$ will fullfil the criterion 
(\ref{chi}). To see whether this is compatible with 
our criterion (\ref{eqn4.22}) we have studied (analytically and numerically) 
the variance, over arbitrary intervals $[v_1^\prime, v_2^\prime]$, 
of the function $N_{\kappa}(v)$. We found that the inequality 
(\ref{eqn4.22}) is satisfied for a constant $C_2^{(\kappa)}$ 
which is of order unity if $\kappa$ stays of order unity. 
[For instance, in the extreme case $\kappa = 1/2$ corresponding 
to a logarithmically divergent incoming energy flux,  
we find $C_2^{(1/2)} \sim 0.2$]. We note, however, 
that if one considers extremely slow decays 
of $N(v)$, i.e. very small exponents $\kappa \rightarrow 0$ 
(completed by a faster decay, or an exact vanishing, before some large  
negative cutoff $v_c$ ), the constant $C^{(\kappa)}_2$ needed 
on the R.H.S. of the variance criterion (\ref{eqn4.22}) tends to zero. 
This signals a limitation of applicability of our simple 
variance criterion based only on the 
quadratic-order approximation $\mu = \mu_2 + \mu_4 + \cdots \simeq 
\mu_2$. Indeed, one can check that, in such an extreme situation 
$\mu_2$ is abnormally cancelled, while $\mu_4$ will be of order unity. 
However, we believe that for generic, non extremely slowly 
varying news functions, the simple ``rule of thumb'' 
(\ref{eqn4.22}) is a reasonable approximation to the (unknown) 
exact collapse criterion. To further check the reasonableness 
of our criterion (\ref{eqn4.16}) we  
turn our attention to some exact solutions. 

\subsection{Exact solutions}
\label{sec3.4}
A first exact, dynamical Einstein-dilaton 
solution is defined  by a news function consisting
of  the simple step function
\beq
N(v) = p \, \theta (v) \, . 
\label{eqn4.18}
\eeq
The corresponding solution  has been independently 
derived by many authors \cite{japanese}. Here, $p$ is a real parameter and the value $|p|=1$ 
defines the threshold for gravitational instability: no singularity 
occurs for $|p| < 
1$, while $|p| \geq 1$ causes the birth of singularity. The metric for this 
solution takes the form (note that $\Omega = 1$) 
\beq
\label{eq5.1}
ds^2 = -du\,dv + r^2\,d \omega^2 \,,
\eeq
with the solutions for  $\phi$, $r$ and  $m$ given as follows: 
\beq
\label{eq5.2}
\phi = 0\,, \quad \quad r = \frac{1}{2}(v-u) \,, \quad \quad m =0\,, 
\quad \quad \mbox{ for $v < 0$}\,,
\eeq
while, for $v \geq 0$,
\bea
\label{eq5.3}
&& \phi(u,v) = - \log \left [ \frac{(1-p)\,v -u}{(1+p)\,v -u} \right ]\,,\\
\label{eq5.4}
&& r^2(u,v) = \frac{1}{4}\,\left [ (1-p^2)\,v^2 - 2v\,u + u^2 \right ] = 
\frac{1}{4}\,\left [(1-p)\,v -u \right ]\, 
\left [(1+p)\,v -u \right ]\,,\\
\label{eq5.5}
&& m(u,v) = -\frac{p^2\,u\,v}{8 r}\,,\\
\label{eq5.6}
&& \mu(u,v) = -\frac{p^2\,u\,v}{\left [ (1-p^2)\,v^2 - 2v\,u +u^2 \right ] }\,.
\eea
Let us note that the perturbation-theory value of $\mu$ for $v > 0$, is 
\beq
\mu_2{(u,v)} = {\rm Var} \, [p \, \theta (x)]_{[u,v]} = - 
\frac{p^2\,u\,v}{(v-u)^2} \, \theta (-u) \, .
\label{eq5.7}
\eeq
The maximum value of $\mu_2$,  reached for  $u = -v$, is $p^2 / 4$. This 
shows that, as expected, the quadratic order criterion (\ref{eqn4.16}) is 
only valid as an order of 
magnitude, but is quantitatively modified by higher-order corrections. This 
example, and a related general theorem of Christodoulou \cite{chi93}, suggest 
that, if the exact criterion were of the type ${\cal S}_2 = C_2$, the constant 
$C_2$ 
should equal $\frac{1}{4}$. (Note that this value is also compatible with the constant 
$C_2^{(1/2)} \sim 0.2$ appropriate to $|v|^{-1/2}$ decay.) 

A second exact solution \cite{CLW}, \cite{TW} is a negative-curvature Friedmann-like homogeneous 
universe. It is defined by the following null data on ${\cal I}^-$ (for $v < 
0$)
\beq
f(v) = - \frac{\sqrt 3}{v} \ , \quad \quad  N(v) = \frac{\sqrt 3}{v^2} \, .
\label{eqn4.19}
\eeq
Here we view this solution as defined by incoming wave data in a flat Minkowski 
background. Actually the data are regular only in some advanced cone $v = T + R 
< 0$, and blow up when $v = 0$. Here, $T$ and $R = \sqrt{X^2 + Y^2 + Z^2}$ are 
usual Minkowski-like coordinates in the asymptotic past. In terms of such 
coordinates the exact solution reads (when $T^2 - R^2 < 1$)
\beq
\label{4.43}
ds^2 = 
\left [ 1 - \frac{1}{(T^2 - R^2)^2} \right] [-dT^2 + dX^2 + dY^2 + dZ^2]\,,
\eeq
\beq
\label{4.44}
\phi = - \sqrt 3 \, \log \left( \frac{T^2 - R^2 -1}{T^2 - R^2 +1} \right) \,.
\eeq
Note that, for simplicity, we have set here $\phi_0 = 0$, and we have also set the 
length scale appearing in $f(v) = - \sqrt 3 \, \ell_0^2 / v$ to $\ell_0 = 1$. 
We shall come back later to the cosmological significance of this solution. Let 
us only note here that, in terms of the null coordinates
\beq
u \equiv T - R \ , \quad \quad v \equiv T + R \, ,
\eeq
the exact mass ratio reads
\beq
\mu(u,v) = \frac{u\,v\,(v-u)^2}{(u^2\,v^2-1)^2}\,,
\eeq
while, starting from (\ref{eqn4.19}), one obtains
\beq
\mu_2{(u,v)} = {\rm Var} (N)_{[u,v]} = \frac{(v-u)^2}{u^3 \, v^3} \, .
\label{eq5.28}
\eeq
One finds that a strength criterion of the form ${\displaystyle \sup_{u \leq 
v}} \ \mu_2 (u,v) = C_2$ is first satisfied (as one increases  $v$ from $-\infty$) 
 when $v = v_{*2}$, with $u = u_{*2}$, where
\beq
v_{*2} = -\ell_0 \left( \frac{4}{27C_2} \right)^{1/4} \ , 
\quad \quad  u_{*2} = 3 v_{*2} \,.
\eeq
For $C_2$ of order 1, this is in qualitative agreement with the exact result that 
the apparent horizon is located at
\beq
\label{eq5.29}
u_{\cal A} = \frac{1}{v^3} \,, \quad \quad \quad 
r_{\cal A} = \frac{1}{2 |v|^3}\,(1 - v^4)^{3/2}\,,
\eeq
so that the apparent horizon and the 
singularity are first ``seen'' at $v_{* \, {\rm exact}} = - \ell_0$.

\section{Transition from the weak-field  to the cosmological  regime}
The following general picture emerges from the previous sections: Let us 
consider as ``in state'' a generic classical string vacuum, which can be 
described as a superposition of incoming wave packets of gravitational 
and dilatonic fields. This ``in state" can be nicely parametrized by three 
asymptotic ingoing, dimensionless news functions $N(v,\theta,\varphi)$, $N_+ 
(v,\theta,\varphi)$, $N_{\times} (v,\theta,\varphi)$.  
 When  all the news stay always significantly below 1, this ``in state'' will 
evolve into a similar trivial ``out state'' made of outgoing wave packets. On the 
other hand, when the news functions reach values of order 1, and more precisely 
when some global measure of the variation \footnote{The argument that the 
collapse criterion should be (at least) invariant under constant shifts of all 
the news functions (corresponding to classically trivial shifts of the 
background fields $g_{\mu \nu}^0$, $\phi_0$) indicates that only the global 
changes of the news matter.} of the news functions, similar to the variance 
(\ref{eqn4.16}), exceeds some critical value of order unity, the ``in state'' will 
become gravitationally unstable during its evolution and  will give 
birth to one or several black holes, i.e. one or several singularities hidden 
behind outgoing null surfaces (event horizons). Seen from the outside of these 
black holes, the ``out" string vacuum will finally look, like the ``in" one, as a 
superposition of outgoing waves.
However, the story is very different if we 
look inside these black holes and shift back to the physically more appropriate 
string conformal frame. First, we note that the structure of black hole 
singularities in Einstein's theory (with matter satisfying $p < \epsilon$) is a 
matter of debate. The work of Belinsky, Khalatnikov and Lifshitz \cite{BKL} has 
suggested the generic appearance of an oscillating space-like singularity. 
However, the consistency of this picture is unclear as the infinitely many 
oscillations keep space being as curved (and ``turbulent'' \cite{Russians},  
\cite{Belinsky}) as time. Happily, the basic gravitational sector of string 
theory is generically consistent with a much simpler picture. Indeed, it has 
been proven long ago by Belinsky and Khalatnikov \cite{BK} that adding a 
massless scalar field (which can be thought of as adding matter with $p = 
\epsilon$ as equation of state) drastically alters the BKL solution by ultimately 
quenching the oscillatory behaviour to end up with a much simpler, monotonic 
approach to a space-like singularity. When described in the string frame, the 
Einstein-frame collapse towards a space-like singularity will represent (if 
$\phi$ grows toward the singularity) a super-inflationary expansion of space. 
The picture is therefore that inside each black hole, the regions near the 
singularity where $\phi$ grows will blister off the initial trivial vacuum as 
many separate pre-Big Bangs. These inflating patches are 
surrounded by non-inflating, or deflating (decreases $\phi$) 
patches, and therefore globally look approximately 
closed Friedmann-Lema\^{\i}tre hot universes. 
This picture is sketched in Fig.~\ref{fig1} and Fig.~\ref{fig2}.
We expect such quasi-closed universes to recollapse 
in a finite, though very long, time (which is 
consistent with the fact that, seen from the outside, 
the black holes therein contained must evaporate in a finite time).
To firm up this picture let us study in detail the 
appearance of the singularity in the simple toy model of Section \ref{sec3}.

\subsection{Negative-curvature Friedmann-dilaton solution}
\label{subsec5.1}
We have already seen that in the toy model there is an infinitely 
extended incoming region where the fields are weak and can be described as 
perturbations upon the incoming background values $(g_{\mu \nu}^0 = \eta_{\mu 
\nu} , \phi_0)$. One expects that the perturbation algorithm described in section \ref{subsec3.1} 
becomes unreliable when corrections become of order unity. In 
particular, one generically expects that the apparent horizon $\mu \equiv 
\frac{2m}{r} = 1$ will roughly divide space-time in two domains: the 
perturbative incoming weak-field domain where $\mu (u,v) \ll 1$, and a 
strong-field domain where $\mu (u,v) \gg 1$. [Actually, we shall see below that 
the precise boundary of the domain of validity of perturbation theory may also 
depend on other quantities than just $\mu$.] This separation in two domains 
is represented in Fig.~\ref{fig3}.

For guidance let us study in detail 
this separation in two domains in the case of the $k = -1$ Friedmann-dilaton 
solution \cite{CLW}, \cite{TW}. 
Let us first write explicitly the solution Eqs. (\ref{4.43}) and (\ref{4.44})
in $(u,v)$ coordinates 
\beq
\label{eq5.18}
ds^2 = -\Omega^2(u,v)\,du \,dv + r^2 \, d \omega^2\,, \quad \quad 
\phi(u,v) = - \sqrt{3}\,\log \left [ 
\frac{u\,v -1}{u\,v +1} \right ]\,,
\eeq
\bea
\label{eq5.25}
&& \Omega^2(u,v) = \left ( 1 - \frac{1}{u^2\,v^2} \right )\,,
\quad \quad r^2(u,v) = \frac{1}{4}\,\left ( 1 - \frac{1}{u^2 \, v^2} \right )\,
(u-v)^2 \,,\\
\label{eq5.26}
&& \mu(u,v) = \frac{u\,v\,(v-u)^2}{(u^2\,v^2-1)^2}\,, 
\quad \quad 
1 - \frac{2m}{r} = \frac{(u^3\,v-1)\,(u\,v^3-1)}{(u^2\,v^2-1)^2}\,.
\label{eq5.27}
\eea
The perturbative algorithm of section \ref{subsec3.1} gives
\bea
\label{eq5.33}
\phi(u,v) &\simeq& \frac{2 \sqrt{3}}{u\,v}\,\left (1 + \frac{1}{3}
\frac{1}{u^2\,v^2} \right ) \,,\\
\label{eq5.34}
r(u,v) & \simeq & \frac{1}{2}\,(v-u)\,\left (1- \frac{1}{2}
\frac{1}{u^2\,v^2} \right )\,,
\eea
which agree with the expansions of Eqs.~(\ref{eq5.25}) and (\ref{eq5.26}). One sees that 
perturbation theory is numerically valid up to, say, $uv \leq \sqrt 2$, at which point 
there is an abrupt transition towards the cosmological singularity  
located at $uv = 1$. Though the transition surface $uv \sim \sqrt 2$ globally 
differs from the apparent horizon $\mu (u,v) =1$, we note that our criterion 
points out to a specific event $(u_{*2} , v_{*2})$ on the apparent horizon 
${\cal A}$ (the point where ${\cal A}$ is first ``seen'' from infinity) which 
lies, roughly, at the intersection of ${\cal A}$ and of the transition surface. 
This confirms that our criterion is able, at least in order of magnitude, to 
correctly pinpoint when and where one should shift from the perturbative regime to 
a different, cosmological-type description. In the case of the solution 
Eqs.~(\ref{eq5.18})--(\ref{eq5.26}),
one sees better the cosmological nature of the strong-field domain by 
introducing the following coordinates
\beq
u = -e^{-\eta + \xi} = T - R \,, \quad \quad  v = - e^{-\eta - \xi} = T +R \,,
\label{eq5.21}
\eeq
satisfying the useful relations
\bea
\label{eq5.22}
&& -T = e^{-\eta}\, \cosh \xi \,, \quad \quad \quad 
R = e^{-\eta}\, \sinh \xi \,,\\
\label{eq5.23}
&& \pa_u = e^{\eta -\xi}\, (\pa_\eta - \pa_\xi) \,,
\quad \quad \pa_v = e^{\eta +\xi}\, (\pa_\eta + \pa_\xi)\,,
\\
\label{eq5.24}
&& e^{-2 \eta} = u\,v\,,\quad \quad \quad \quad \quad \quad e^{2 \xi} = \frac{u}{v}\,.
\eea
In terms of the coordinates $(\eta , \xi)$ the solution reads \cite{CLW}, \cite{TW}
\bea
\label{eq5.16}
&& ds^2 = a^2(\eta) \left [ -d \eta^2 + d \xi^2 + \sinh^2 \xi\,d \omega^2 \right ]\,,\\
\label{eq5.17}
&& a^2(\eta) = 4\cosh \eta\,\sinh(-\eta)\,, \quad \quad 
e^{-\phi} = \left (-\frac{\sinh \eta}{\cosh \eta} \right )^{\sqrt{3}}\,.
\eea
We note that the solution is regular in the domain $-\infty < \eta < 0$, $0 
\leq \xi < +\infty$ (which corresponds to the past of the hyperboloid $T^2 - 
R^2 = 1$ in Minkowski-like coordinates) and that a space-like singularity is 
reached at $\eta = 0$ (i.e. $uv = T^2 - R^2 = 1$). Let us also note the 
expressions
\bea
r^2 (\xi , \eta) = a^2 (\eta) \sinh^2 \xi = 4 \cosh \eta \sinh (-\eta) \sinh^2 \xi \, ,
\eea
\beq
\label{eq5.30}
1 - \frac{2 m}{r} = \sinh^2 \xi\, \left [ \coth^2 \xi - \coth^2 2 \eta \right ] \,,
\eeq
telling us that the apparent horizon is located at $2\eta = -\xi$. Near the 
singularity, $\eta \rightarrow 0^-$,
 the dilaton blows up logarithmically while the metric coefficients 
have the following power-law behaviours:
\beq
\phi \sim - \sqrt 3 \, \log (-\eta)\,, \quad \quad 
\Omega^2 (\xi , \eta) \equiv a^2 (\eta) \sim -4 \eta\,, 
\quad \quad r^2 (\xi , \eta) \sim -4 \eta \sinh^2 \xi \, .
\eeq
Before leaving this example we wish to emphasize some features of it which
appear to follow from its homogeneity and could be misleading for the general case. 
The singular boundary  terminates, in this case, on ${\cal I}^-$, 
instead of on ${\cal I}^+$, as one generically expects.
 The apparent reason for this is the
singularity in the flux at $v=0$ which creates a future boundary on ${\cal I}^-$.
We believe that, in this case, a better description of physics is obtained by going to
new non-Minkowskian coordinates of Milne's type (see \cite{V97}) which automatically
incorporate the singularity on ${\cal I}^-$. However, 
if one restricts oneself to more regular initial data,  having
a finite integrated energy flux (generalized pulse-like data), the singular boundary
should never come back to ${\cal I}^-$. [At most,  it could end at space-like
infinity ${i}^0$, for non integrable total energy flux].

\subsection{Kasner-like behaviour of Einstein-dilaton singularities}
\label{subsec5.2}
\begin{figure}
\centerline{\epsfig{file=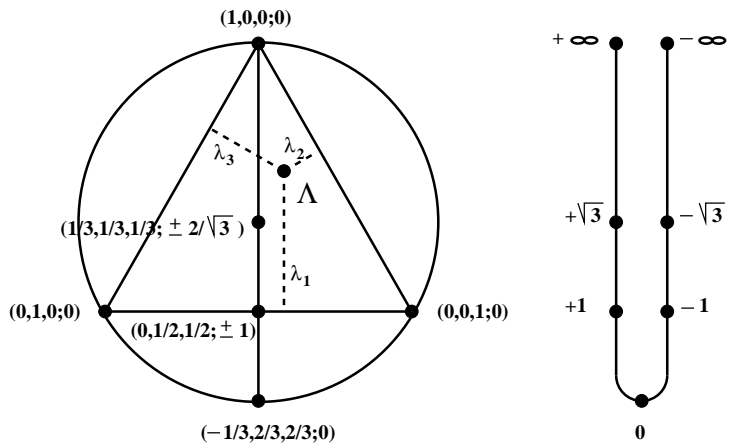,width=0.35\textwidth,angle=-90}}
\vskip 0.5truecm
\caption{\sl Geometric representation of the Einstein-frame 
Kasner exponents $(\lambda_1, \lambda_2, \lambda_3; \gamma)$ 
of Einstein-dilaton cosmological singularities. The three $\lambda$'s, 
such that $\sum_a \lambda_a =1$ are the orthogonal distances 
to the three sides of an equilateral triangle. The constraint 
$\gamma = \pm \sqrt{2}\, \sqrt{1 - \sum_a \lambda^2_a}$ restricts the representative $\Lambda$ of $\lambda_a$ to stay on a two-sided disk circumscribed 
around the triangle. The pure-Einstein cases ($\gamma = 0$) correspond 
to the circumscribing circle. In the spherically symmetric case 
($ \lambda_2 = \lambda_3$), $\Lambda$ runs over a bissectrix of the triangle. 
The basic parameter $\alpha$ of Eqs.~(\ref{phi}) and (\ref{lam}) 
runs over a full real line (shown folded on the right of the figure 
) and is mapped to the bissectrix via horizontal lines.}
\label{fig4}
\end{figure}

Coming back to the generic case of an arbitrary Einstein-dilaton cosmological 
singularity, previous works \cite{V97} have shown that, in a suitable synchronous 
coordinate system (Gauss coordinates), the asymptotic behaviour of the string frame 
metric near the singularity reads (in any space-time dimension $D$)
\beq
\label{strkasn1}
ds_S^2 \sim -dt_S^2 + \sum_{a=1}^{D-1} (-t_S)^{2\alpha_a (x)} 
(E_i^a (x) \, dx^i)^2 \, ,
\eeq
\beq
\label{strkasn2}
\phi (x,t) = \phi (x,0) + \sigma (x) \log (-t_S) \, ,
\eeq
where $E_i^a (x)$ is some $(D-1)$-bein and where ($1 \leq a \leq D-1$)
\beq
\label{strrel}
\sum_{a=1}^{D-1} \alpha_a^2 = 1 \,, \quad \quad 
\sigma = \left( \sum_{a=1}^{D-1} \alpha_a \right) - 1 \,.
\eeq
In the Einstein frame this asymptotic behaviour reads
\beq
\label{kasn1}
ds_E^2 \sim - dt_E^2 + 
\sum_a (-t_E)^{2 \lambda_a (x)} (e_i^a (x) \, dx^i)^2 \,,
\eeq
\beq
\label{kasn2}
\phi (x,t) = \phi (x,0) + \gamma (x) \log (-t_E) \,,
\eeq
where $e_i^a (x)$ is proportional to $E_i^a (x)$ and where
\beq
\sum_{a=1}^{D-1} \lambda_a = 1 \,, \quad \quad 
\gamma = \pm \sqrt 2 \ \sqrt{1 - \sum_{a=1}^{D-1} \lambda_a^2} \,.
\label{cons}
\eeq
The ``Kasner'' exponents, $\alpha_a$ in the string frame or $\lambda_a$ in the 
Einstein frame, can vary continuously along the singularity. The string 
parametrization is the most global one as it shows that $\alpha_a$ runs freely 
over a unit sphere in $\Rb^{D-1}$ while the exponent for $e^{\phi}$ is a linear 
function of the ``vector'' $\alpha_a$. In the Einstein frame, the exponents 
$\lambda_a$ are restricted by the linear equality $\sum_a \lambda_a = 1$ and the 
quadratic constraint $\sum_a \lambda_a^2 \leq 1$. A convenient geometric 
representation of these constraints, when $D=4$, is   
to consider (in analogy with Mandelstam's  
variables $s,t,u$) that the 3 $\lambda_a$'s represent, in some Euclidean plane, 
the orthogonal distances of a point $\Lambda$ from the three sides of an 
equilateral triangle (counted positively when $\Lambda$ is {\it inside} the 
triangle). The quadratic constraint then means that $\Lambda$ is restricted to 
stay inside the circle circumscribing the triangle. The sign ambiguity in 
Eq.~(\ref{cons}) means that the parameter space is in fact a {\it two-sided disk}, namely 
the two faces of a coin circumscribed around the triangle. See Fig.~\ref{fig4}.

\begin{figure}
\vskip -1truecm
\centerline{\epsfig{file=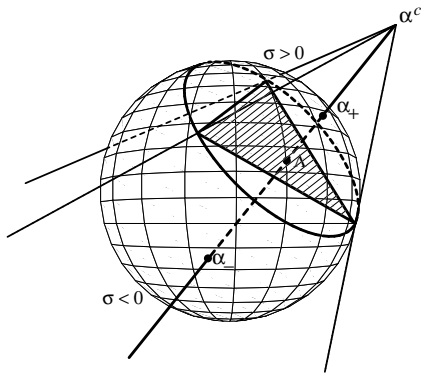,width=0.36\textwidth,angle=0}}
\vskip -1.5truecm
\caption{\sl Geometrical representation of the link 
between the two-sided disk of Fig. \ref{fig4} on which 
the Einstein-frame Kasner exponents $\lambda_a$ live, 
and the sphere, $\sum_a \alpha^2_a=1$, on which the string-frame
Kasner exponents $(\alpha_1, \alpha_2, \alpha_3)$ live. The map $
\alpha_a \rightarrow \lambda_a$ is a stereographic projection, 
from the center $(\alpha_a^c)= (1,1,1)$. The upper side of the disk 
comes from the projection 
of the polar 
cap $\sigma \equiv (\sum_a \alpha_a) -1 > 0$ located on the same side as 
$\alpha^c$, while the lower side of the disk comes from 
projection of the antipolar 
cap $\sigma < 0$. The points $\alpha_+$ on the polar cap, and 
$\alpha_-$ on the antipolar cap, are both projected on the same point 
$\Lambda$ of the disk. The tangents to the sphere issued from $\alpha^c$ 
touch the sphere along the circle limiting the disk (on which $\sigma=0$).}
\label{fig5}
\end{figure}

The link between the 
two seemingly very different parameter spaces (the string-frame sphere 
parametrized by the $\alpha$'s and the Einstein-frame two-sided disk 
parametrized by the $\lambda$'s) is geometrically very simple: one obtains the 
disk by a stereographic projection of the sphere, from the projection 
center $(\alpha_a^c) 
= (1,1,1)$ (outside the sphere) onto the plane dual to this center (i.e. the 
plane spanning the circle of tangency to the sphere of the straight lines issued 
from $\alpha^c$). Algebraically (in any-space-time dimension $D$)
\beq
\lambda_a = \frac{\alpha_a(D-2) - {\sigma}}{D-2-{\sigma}}\,,
\quad \quad 
\gamma = \frac{\sqrt{ 2\,(D-2)}\,\sigma}{D-2 - {\sigma}}\,,
\eeq
where $\sigma \equiv (\sum_a \alpha_a) - 1$. Note that there is no sign ambiguity 
in the map $\alpha_a \rightarrow \lambda_a$. Coming back to the case $D = 4$, as 
$\alpha_a$ runs over a unit sphere in $\Rb^3$, $\lambda_a$ covers twice the disk 
circumbscribed to an equilateral triangle. Each side of the disk is the image, 
via the stereographic projection, of a polar ($\sigma > 0$) or antipolar 
($\sigma < 0$) ``cap'' of the sphere. This geometrical picture of the link 
between the $\alpha$-sphere and the two-sided $\lambda$-disk in represented in 
Fig.~\ref{fig5}. 
It is useful to keep in mind this 
geometrical picture because it shows clearly that one can continuously pass from 
one side of the disk to the other, i.e. the sign of the Einstein-frame 
``Kasner'' exponent of $e^{\phi} \sim (-t_E)^{\gamma (x)}$ can change as one moves 
along the singularity. Physically, this means a very striking inhomogeneity near 
the singularity: the patches where $\gamma (x) > 0$ (decreasing $g^2 = 
e^{\phi}$) will shrink (in string units) while only the patches where $\gamma 
(x) < 0$ (growing $g^2$) can represent pre-Big Bangs. Finally, we note 
that Belinsky and Khalatnikov \cite{BK} have proven that, in the generic case, the 
Einstein-frame Kasner exponents $(\lambda_1, \lambda_2, \lambda_3)$ 
must all be positive for the ultimate stable asymptotic approach 
to the singularity. This corresponds to $\Lambda$ being inside the triangle
of Fig.~\ref{fig4}, i.e. in the hatched region of Fig.~\ref{fig5}.
However, this generic restriction does not apply 
in the special case of our toy model as we explain below.
\subsection{Behaviour near the singularity in inhomogeneous, 
spherically symmetric solutions} 
\label{subsec5.3}
Let us now restrict ourselves to our toy model and study its cosmological-like 
behaviour near the singularity. First, we note that spherical symmetry will 
impose that two of the metric Kasner exponents (those corresponding to the 
$\theta$ and $\varphi$ directions) must be equal, say $\lambda_2 = \lambda_3$. 
Geometrically, this means that the Einstein-frame Kasner exponents run only over 
the intersection of the above two-sided disk with a bissectrix of the triangle.
Second, we note that because of the vanishing, in spherical symmetry, of the crucial 
dreibein connection coefficient $\vec{e}_1 \cdot (\nabla \times \vec{e}_1)$, 
generically responsible for causing the expansion in the (radial) direction $1$ 
to oscillate as $t \rightarrow 0^-$ \cite{BKL}, a negative value 
of $\lambda_1$, corresponding to an expansion in the radial direction, is allowed 
as ultimate asymptotic behaviour. As a consequence, even the portion of 
the bissectrix outside the triangle is allowed as asymptotic state. We shall prove directly
this fact below by constructing a consistent expansion near the singularity 
for $\lambda_1$ of any sign. 

The mass-evolution equations (\ref{eq1.8}), (\ref{eq1.9}), together with 
Eqs.~(\ref{rel1}), (\ref{rel2}) show that, starting from 
some regular event at the center, $r=0$, where $m=0$, the mass $m (u,v)$ will 
grow if we follow a  outgoing characteristic (i.e. a null 
geodesic) $u =$ const. If this characteristic crosses the apparent horizon 
${\cal A}$, Eqs.~(\ref{rel1}), (\ref{rel2}) then show that $m(u,v)$ 
continues to grow within the trapped region ${\cal T} (2m/r > 1)$ if one moves 
along {\it either} (future) outgoing or ingoing characteristics. This means that, 
within ${\cal T}$, $m(u,v)$ grows in all future directions. As the space-like 
singularity \footnote{We consider here only the 
``non central'' singularity, i.e. 
the part of the singularity which can be reached by future outgoing 
characteristics issued from the regular center $r=0$. As shown by \cite{chi93}
there exists also, at the ``intersection'' between the regular center and 
${\cal B}$, a singularity reachable only via ingoing characteristics. 
Note that $r(u,v)$ tends to zero both at the center 
and at the singularity (central or 
non-central).} ${\cal B}$ lies in the future of ${\cal T}$, we see that the mass 
function will necessarily grow toward ${\cal B}$, and therefore, either will 
tend to a finite limit on ${\cal B}$, or it will tend monotonically to $+\infty$ as 
$r \rightarrow 0$ with fixed $v$ (or $u$). Motivated by the Kasner-like 
behaviours Eqs.~(\ref{kasn1}) and (\ref{kasn2})
 we expect the following generic asymptotic behaviour of $m$ 
on the singularity (here written in $(v,r)$ coordinates)
\beq
\label{mass}
m(v,r) \sim C_m(v) \, \left ({r\over \ell_0} 
\right )^{-\alpha^2 (v)} \ \hbox{as} \ r 
\rightarrow 0 \,, v \,\,\hbox{fixed}
\eeq
with some real $v$-dependent exponent $\alpha^2 (v)$. [Here $\ell_0$ 
denotes some convenient length scale.]
It follows that $\mu = 2m/r$ always tends to $+\infty$ on ${\cal B}$ 
(even when $m$ tends to a finite limit). This suggests that we can 
consistently compute the structure of the fields near ${\cal B}$ by using an 
``anti-weak-field'' approximation scheme in which $\mu \gg 1$ (instead of the 
weak-field algorithm of section~\ref{subsec3.1} 
where we used $\mu \ll 1$). The expected domain of validity of such 
an ``anti-weak-field'', or ``cosmological-like'', expansion 
is sketched in Fig.~\ref{fig3}.

To leading order 
in this large-$\mu$ limit Eq.~(\ref{eq1.19}) yields
\beq
\label{eq4.11}
\frac{1}{4}\left ( \frac{ \pa \phi}{\pa \log r} \right )^2_{v} = 
- \left (\frac{\partial \log m}{\pa \log r} \right )_{v}\,,
\eeq
from which we obtain the following asymptotic behaviour of $\phi$ on ${\cal 
B}$
\beq
\label{phi}
\phi (r,v) \sim 2 \alpha (v) \log r + C_\phi (v) \, .
\eeq
We see now that the basic Kasner-like-exponent on ${\cal B}$ is half the 
coefficient of $\log r$ in $\phi (r,v)$ as $r \rightarrow 0$, with fixed $v$. 
The Kasner-like exponent of $m(v,r)$ is the square of that basic exponent. 
Moreover, the function $C_m(v)$ appearing in the asymptotic behaviour 
of $m(v,r)$ on the singularity is not independent from the functions 
$\alpha(v)$ and $C_\phi(v)$. Indeed, to leading order Eq.~(\ref{new}) 
reads
\beq
\frac{\pa \log m}{\pa v} = - \frac{1}{2}\,\frac{\pa \phi }{\pa \log r}\,
\frac{\pa \phi}{\pa v}\,.
\eeq
This relation is identically satisfied at order $r^0\,\log r$, and yields at order $r^0$ 
\beq
\frac{\pa \log C_m}{\pa v} = - \alpha(v)\,\frac{\pa C_\phi}{\pa v}\,.
\label{5.26}
\eeq
Using now Eq.~(\ref{eq1.18}), written as
\beq
\left( \frac{\partial \beta}{\partial \log r} \right)_v = 
\left( \frac{1}{2} \, 
\frac{\partial \phi}{\partial \log r} \right)^2\,,
\eeq
we further get the asymptotic behaviour of $\beta (v,r)$
\beq
\label{beta}
\beta (v,r) \sim \alpha^2 (v) \log r + C_\beta (v) \, .
\eeq
Finally, it is easy to see that the asymptotic behaviour  of the 
metric on ${\cal B}$ in $(v,r)$ coordinates
\beq
ds^2 \sim \frac{2 C_m (v)}{r} \, 
e^{2C_\beta (v)} \, \left ({r\over \ell_0}\right )^{\alpha^2 (v)} 
\, dv^2 + 2 e^{C_\beta (v)} \, \left ({r\over \ell_0}
\right )^{\alpha^2 (v)} \, dv \, dr + r^2 \, d\omega^2\,,
\eeq
is indeed of the expected Kasner form (\ref{kasn1}) by introducing the 
cosmological time 
\beq
(-t) \sim r\,\left ({r\over \ell_0}
\right )^{\frac{1}{2} (\alpha^2 (v) + 1)}\,.
\eeq
The transformed metric reads, to dominant order as $r \rightarrow 0$, i.e. $t 
\rightarrow 0^-$,
\bea
ds^2 \sim&& -dt^2 + C_{\rho} (v) \, (-t)^{2\lambda_1 (v)} \, \left (
dv + \frac{r\,e^{-C_{\beta}}\,dr}{2 C_m} \right )^2  \nonumber \\
&& + C_{\omega} (v) \, (-t)^{2\lambda_2 (v)} \, [d\theta^2 + \sin^2 \theta \, d \varphi^2]\,, 
\eea
\beq
\phi \sim \gamma (v) \log (-t) \,,
\eeq
\beq
\label{lam}
\lambda_1 (v) = \frac{\alpha^2 (v) - 1}{\alpha^2 (v) + 3} \,,
\quad \quad 
\lambda_2 (v) =  \frac{2}{\alpha^2 (v) + 3} \,,
\quad \quad 
\gamma (v) = \frac{4 \alpha (v)}{\alpha^2 (v) + 3} \,.
\eeq
The basic parameter $\alpha (v)$ runs over the real line $- \infty < \alpha < 
+\infty$. One easily checks that the relations (\ref{cons}) are satisfied. 
Some cases having a particular significance are illustrated in Fig.~\ref{fig4}: 
When $\alpha = 0$, 
one gets $(\lambda_1 , \lambda_2 , 
\lambda_3) = \left( - \frac{1}{3}, \frac{2}{3}, \frac{2}{3} \right)$ and $\gamma 
= 0$ which describes a Schwarzschild-like singularity (when viewed with a 
cosmological eye). 
When $\alpha = \pm 1$, 
one gets $(\lambda_1 , \lambda_2 , \lambda_3) = \left( 0, \frac{1}{2}, 
\frac{1}{2} \right)$ and $\gamma = \pm 1$ which is locally similar to the 
scale-invariant solution \cite{japanese} also described earlier. 
When $\alpha = \pm \sqrt{3}$, one gets $(\lambda_1 , 
\lambda_2 , \lambda_3) = \left( \frac{1}{3}, \frac{1}{3}, \frac{1}{3} \right)$ 
and $\gamma = \pm 2 / \sqrt 3$ which describes an isotropic cosmological 
singularity of the type of the exact solution \cite{CLW}, \cite{TW} we discussed above. 
As illustrated in Fig.~\ref{fig4},  
when $\alpha$ runs over $\Rb$ its image, defined by Eq.~(\ref{lam}), 
runs 
twice over the bissectrix of the Mandelstam triangle (intersected with the 
circumbscribed disk). The only pure-Einstein case ($\gamma = 
0$) reached at finite $\alpha$ is the Schwarzschild-like case $\alpha 
= 0$. In principle, another pure-Einstein behaviour occurs  when $\vert 
\alpha \vert \rightarrow \infty$, leading to $(\lambda_1 , \lambda_2 , 
\lambda_3) = (1,0,0)$ i.e. $ds^2 \sim -dt^2 + t^2 \, d\rho^2 + d \omega^2 = - d 
\tau^2 + d \xi^2 + d\theta^2 + \sin^2 \theta \, d\varphi^2$. This would be an 
Einstein-cylinder-like universe. We believe, however, that $\vert \alpha (v) 
\vert$ will stay bounded $(\vert \alpha (v) \vert < \alpha_M < + \infty)$ as $v$ 
runs over ${\cal B}$ and that the only Einstein-like case will be crossed when 
$\alpha (v)$ changes along ${\cal B}$.

We have confirmed our asymptotic analysis of the structure of the singularity 
(based on the Ans\"atze (\ref{mass}), (\ref{phi}) and (\ref{beta})
in two ways. First, we have verified that it is 
consistent with the partial, but rigorous, results given in Ref.~\cite{chi93}. 
Namely, in $(u,v)$ coordinates, our Ans\"atze are compatible with the result 
that the function $r^2 (u,v)$ is $C^1$ on ${\cal B}$ if we posit that the 
conformal factor behaves like $\Omega^2 \sim r^{\alpha^2 (v) - 1}$ (while 
$m \sim r^{-\alpha^2 (v)}$). A stronger check of our assumptions is obtained by 
going beyond the leading order approximation, thus extending some
results of \cite{V97}. In analogy to section \ref{subsec3.1}, where we 
set up a complete perturbation algorithm in the weak-field domain, we have shown 
that, starting from the leading-order terms Eqs.~(\ref{mass}), (\ref{phi}) 
and (\ref{beta}), containing 3 arbitrary ``seed 
functions'', $\alpha (v)$, $C_\phi (v)$, $C_\beta (v)$ [from which 
$C_m(v)$ can be determined using Eq.~(\ref{new})], it is 
possible to set-up an all-order iterative scheme generating a formal solution of 
the spherically symmetric Einstein-dilaton system in the strong-field, 
cosmological-like domain. Note that, $C_\beta(v)$ being a pure gauge function 
(it suffices to introduce $v^\prime = \int e^{C_\beta(v)} \, dv$ 
to gauge $C_\beta(v)$ away), the formal solution thus generated contains two 
physically arbitrary 
functions of $v$, which is indeed the freedom that should be present 
in a generic solution \footnote{Here, contrary to what happened above, the variable $v$ varies 
only on a half-line, $[v_*, +\infty]$ where $v_*$ marks the birth of ${\cal B}$. 
As was said above two functions on a half-line (or one function on the full line) 
is the correct genericity in our toy model.}.
Details of this scheme are given in Appendix \ref{app}. We 
only mention here the general structure of the expansions so generated near the 
singularity, i.e. as $r \rightarrow 0$, with fixed $v$: (see Eq.~(\ref{ulapp}) in the appendix)
\bea
\label{appen1}
&& \phi(v,r) = \phi_0(v,r) +
\sum_{n,m,l} a_{n m l}^{(\phi)}\,r^{n(\alpha^2(v) +1)}\,
r^{2 m}\,P_l^{(\phi)}(\log r)\,,\\
\label{appen2}
&& \beta(v,r) = \beta_0(v,r) +
\sum_{n,m,l} a_{n m l}^{(\beta)}\,r^{n(\alpha^2(v) +1)}\,
r^{2 m}\,P_l^{(\beta)}(\log r)\,,\\
\label{appen3}
&& m(v,r) = m_0(v,r) +
\sum_{n,m,l} a_{n m l}^{({\rm m})}\,r^{n(\alpha^2(v) +1)}\,
r^{2 m}\,P_l^{({\rm m})}(\log r)\,.
\eea
Here $n,m,l$ are integers with $l \leq m$, $a_{n m l }^{(\psi)}$ 
and the coefficients of the polynomials $P_l^{(\psi)}$ are functions of $v$,  
and $(\phi_0, \beta_0, m_0)$ are the leading-order terms given 
by Eqs.~(\ref{mass}), (\ref{phi}) and (\ref{beta}).
This expansion actually contains two intertwined series: 
an expansion in powers of $x = r^{1+ \alpha^2}$ and a more complicated 
series in $r^{2 m}\,(\log r)^l$. The second expansion 
is linked to the $v$-gradients of the seed functions $\alpha(v)$,  
$C_\phi(v)$, $C_\beta(v)$, while the first expansion is present even in the simple 
``homogeneous'' case where $\alpha$, $C_\phi$ and $C_\beta$ do not depend on $v$. 
The fact, exhibited by Eqs.~(\ref{appen1})--(\ref{appen3}), that the expansion in 
the homogeneous case proceeds along powers of $r^{\alpha^2+1}$ confirms a recent 
conjecture by Burko \cite{Burko}.

\subsection{Exact homogeneous cosmological solutions}
\label{subsec5.4}
Let us consider in more detail the ``homogeneous'' case where 
the seed functions $\alpha$, $C_\phi$ and $C_\beta$ have no spatial 
variation along the singularity.
As it turns out, it is possible to resum exactly all the terms 
of the ``homogeneous'' expansion. Indeed, the analog of the Schwarzschild solution 
($t$-independent, spherically symmetric solution) for the Einstein-dilaton 
system has been worked out analytically long ago by 
Just \cite{Just} (see also \cite{others}), using the special gauge where 
$g_{00}\,g_{rr} =-1$. Contrary to the case of the Schwarzschild 
solution, this solution 
cannot be continuously extended (through a regular horizon) down to a cosmological 
singularity 
at $r=0$. However, it is easy to see that the following cosmological-like
background (which is related to Just's original solution by formally 
extending the radial variable $\overline{r} = l\,x$ ``below'' the curvature
singularity at $\overline{r} = a$) is still a solution 
of the Einstein-dilaton system: 
\bea
\label{ap1}
d s^2 &=& l^2 \left [ \left (\frac{1-x}{x} \right )^b\,d t^2 - 
\left (\frac{x}{1-x} \right )^b\,d x^2 + x^{1 +b}\,(1-x)^{1-b}\, d \omega^2 \right ]\,,\\
\label{ap2}
\phi &=& \pm \sqrt{1 -b^2}\,\log \left ( \frac{1-x}{x} \right )\,.
\eea
Here the (formerly radial) variable $x$ varies between 
$0$ and $1$ and is ``time-like'', while the (formerly time) variable $t$ 
is ``space-like''. The cosmological universe evolves from 
a Big Bang at $x=0$ to a Big Crunch at $x =1$. 
It has two arbitrary parameters, a scale parameter $l$, and the dimensionless
$b$, $ -1 \leq b \leq 1$. Near the singularity 
at $x=0$, the link between $b$ and the parametrization 
used above is 
\beq
\alpha = \mp \sqrt{\frac{1-b}{1+b}}\,, \quad \quad b = \frac{1 - \alpha^2}{
1 + \alpha^2}\,.
\eeq
Note that the behaviour near the other singularity at $x=1$ 
is obtained by changing the sign of $b$ and by changing 
$\alpha$ into $ -1/\alpha$. Some cases of this homogeneous 
solution are of special significance: $b=1$ is Schwarzschild (with $x=r/2m$), 
$b=0$ belongs to the class of scale-invariant solutions
\cite{japanese} discussed above and $b=-1/2$ interpolates 
between a locally isotropic cosmological solution 
($\alpha = \mp \sqrt{3}$) at $x=0$ and an anisotropic one 
($\alpha = \pm 1/\sqrt{3}$) at $x=1$. Note that, in spite of the homogeneity and isotropy
at $x=0$, the latter special solution differs from the (everywhere) 
homogeneous-isotropic solution described earlier.

\subsection{Discussion}
\label{subsec5.5}
Ideally, the two formal expansions we have constructed, the weak-field one 
Eqs.~(\ref{eqn4.1a})--(\ref{eqn4.1c}), 
and the strong-field one Eqs.~(\ref{appen1})--(\ref{appen3}), 
should match at some intermediate 
hypersurface, like the apparent horizon $2m/r = 1$ which looks like a natural 
borderline between the two near domains. If this matching were analytically doable, 
it would determine all the (so-far) arbitrary ``seed functions'' of the 
strong-field scheme in terms of the unique arbitrary function of the weak-field 
one, namely the asymptotic waveform $f(v)$ (or, equivalently, the news $N(v) = 
f'(v)$). However, it is clearly too naive to expect to perform this matching 
perturbatively: neither of the two expansions is expected to be 
convergent \footnote{The possible presence of a finite number of BKL oscillations 
\cite{BK} near ${\cal B}$ suggests that the formal strong-field expansion has 
very bad convergence properties.} (they are probably only asymptotic). Even if 
they converge on some domain they probably both break down before reaching a 
possible overlap region where they might be matched. At this stage, we can only 
state that, in principle, all the seed functions of the strong-field scheme are 
some complicated non linear and  non local functionals of $N(v)$. It would be particularly 
interesting to study the functional dependence of the Kasner exponent 
$\alpha (v)$ on $N(v)$. By causality (i.e. a domain of dependence argument) we 
know that $\alpha (v)$, at advanced time $v$, depends only on $N(v')$ on the 
interval $v' \leq v$. When starting from a generic $N(v)$ ``of order unity'', we 
expect that the resulting $\alpha (v)$ will also be of order unity. A physically 
very important issue is the sign of $\alpha (v)$, i.e. the sign of $\gamma (v)$. 
Indeed $\alpha > 0$ means a decreasing $\phi$, while $\alpha < 0$ means that 
$\phi$ grows near ${\cal B}$. Let us note that, to lowest-order of weak-field 
perturbation theory, the value of $\phi (v,r)$ at the (regular) center ($r=0$) 
is
\beq
\lim_{r \rightarrow 0} \phi_1 (v,r) = \lim_{r \rightarrow 0} \left[ \phi_i + 
\frac{f(v) - f(v-2r)}{r} \right] = \phi_i + 2 f' (v) = \phi_i + 2 N(v) \, .
\label{eqn5.18}
\eeq
Here $\phi_i$ denotes the background value at past infinity. From this result we 
expect that, if $N(v)$ is a simple Gaussian-like wave packet, 
$N(v) = A \exp \left [- (v 
- v_i)^2 / \ell_i^2\right ]$, with a large enough dimensionless amplitude for leading to 
collapse, i.e. $\vert A \vert \gtrsim 1$, the local values of $\phi$ near the 
collapsing central region will, at first, grow if $A > 0$, and decrease if $A < 
0$. In this simple case, we therefore expect $\alpha (v)$ to have the opposite 
sign of $A$ near the ``central'' region of ${\cal B}$, i.e. for $v \lesssim v_i$ 
where the gravitational instability sets in. But, further away on ${\cal B}$, 
i.e. for $v \gg v_i$, the sign of $\alpha (v)$ can change. We also expect that, 
as $v \rightarrow + \infty$, $\alpha (v)$ will tend to zero, corresponding to a 
Schwarzschild-like singularity (in the case of well localized incoming $N(v)$ 
packets). This conjectured link between the sign of $\alpha (v)$ and the sign of 
$N(v)$ is confirmed by the exact isotropic solution Eq.~(\ref{eq5.18})
(with a positive growing 
$N(v) = \sqrt 3 / v^2$, and $\alpha = - \sqrt 3$), 
as well as by the scale invariant solution Eq.~(\ref{eq5.3}). 
Some numerical calculations \cite{Burko} seem to 
confirm the general picture we propose (in particular the interesting 
possibility that the sign of $\alpha (v)$ changes several times on ${\cal B}$ as 
$v$ varies, before reaching a Schwarzschild-like asymptotic regime $\vert \alpha 
(v) \vert \rightarrow 0$ as $v \rightarrow +\infty$). We plan to study in more 
detail these issues in a future publication.

\section{A Bayesian look at  pre-Big Bang's ``fine-tuning"}
\label{sec6}
{} From its inception \cite{Veneziano} it was pointed out that 
a successful pre-Big Bang 
scenario must rely on a ``reservoir'' of inflationary $e$-folds 
during its perturbative phase. This is given by two small numbers, the initial
curvature scale in string units, $H_i \ell_s$, and the
initial string coupling constant $g_i^2 = e^{\phi_i} \ll \!\!< 1$. [In this section, 
the index $i$ is used for labelling ``initial'' quantities.] The need of large 
(or small) numbers  has been recently discussed at length \cite{TW}, 
\cite{KLB} and used to criticize the naturalness of the PBB scenario. In particular,
it was pointed out 
\cite{TW}, \cite{KLB}  that, as soon as one goes beyond the simple 
spatially-flat, homogeneous cosmology framework, the total duration
of the perturbative dilaton-driven phase is finite so that
the resolution of the homogeneity/flatness problems requires
\beq
g_i = e^{\phi_{i}/2} < 10^{-26} \,, \quad \quad 
\frac{L_i}{\ell_s} \gtrsim \frac{M_s}{H_i} > 10^{19} \,.
\label{eq6.1}
\eeq
Here, $L_i$ denotes the spatial homogeneity scale of the  PBB universe at the beginning
of its inflationary phase, 
and $H_i \gtrsim L_i^{-1}$ its time-curvature scale (Hubble parameter).
 
In this Section we shall systematically work with string-frame 
quantities, even when we refer to results 
discussed in previous sections in the Einstein frame.
 For instance, the time-scale  $\ell_i$, characterizing the rate of variation of the 
news functions around the advanced time $v_i$ and leading to 
gravitational instability, is now (locally) measured in string units. In any case, we are
essentially working with dimensionless ratios which are unit-independent: 
$(\ell_i / \ell_s)_S = (\ell_i / \ell_s)_E = g_i (\ell_i / \ell_{\rm 
Planck})_E$. We wish  to emphasize here that, by combining our 
stochastic-like instability picture with a Bayesian 
approach to the {\it a posteriori} probability of being in a position of asking 
fine-tuning questions, the issue of the naturalness of the PBB scenario is 
drastically changed. By ``Bayesian approach'' we mean taking into account 
the selection effect that fine-tuning questions presuppose the existence of a 
scientific civilization. As emphasized long ago by Dicke \cite{D61} and Carter 
\cite{C74}, civilization-related selection effects can completely change the 
significance of large numbers or of apparent coincidences. 
Linde and collaborators explored several aspects of the Dicke-Carter ``anthropic 
principle'' within the inflationary paradigm, and emphasized the necessity 
to weigh {\it a posteriori} probabilities by the physical volume 
of inflationary patches \cite{lindebook2}.
We shall here follow 
Vilenkin \cite{V95}, and his ``principle of mediocrity'', according to which the 
unnormalized {\it a posteriori} probability of a random scientific civilization 
to observe any values of the PBB parameters $g_i$, $L_i$ and $H_i$ is obtained 
by multiplying the corresponding {\it a priori} probability by the number of 
civilizations (over the whole of space and time) associated with the values $g_i$, 
$L_i$ and $H_i$: ${\cal N}_{civ} (g_i , L_i , H_i)$.

\subsection{A priori and a posteriori probability distributions for $g_i$ 
and $H_i$}
\label{subsec6.1}
 This approach can be 
applied to our case if we think of the initial past-trivial string vacuum as 
made of a more or less stochastic superposition of incoming waves (described by 
complicated news functions $N(v,\theta ,\varphi)$, $N_+ (v,\theta ,\varphi)$, 
$N_{\times} (v,\theta ,\varphi)$ having many bumps, troughs and ramps). This 
stochastic bath of incoming waves will generate a rough sea of dilatonic and 
gravitational fields. If the input dimensionless wave forms can 
reach values of order unity, we expect that the local 
conditions for gravitational instability will be satisfied 
 at several places in space and time. This will give birth 
to an ensemble of bubbling baby universes, with a more or less random 
distribution of initial parameters $g_i$ and $H_i$, and with initial spatial 
homogeneity scales 
$L_i \gtrsim H_i^{-1}$. Indeed, the analysis of the previous 
sections has shown (in our toy model) that gravitational 
instability will set in 
on a spatio-temporal scale $\sim \ell_i$ when a rising wave of news function 
grows by ${\cal O} (1)$ on an advanced-time scale $\ell_i$. From our variance 
criterion Eq.~(\ref{eqn4.16}), and the rough validity of weak-field perturbation
 theory nearly until 
the sharp transition to a cosmological-type behaviour, the work of the previous 
sections has shown that the initial, advanced-time scale $\ell_i$ is propagated 
via ingoing characteristics with little deformation down to the strong-field 
domain (see Fig.~\ref{fig3}), where it appears as a spatial homogeneity scale, i.e.
\beq
L_i \sim \ell_i \, .
\label{eq6.2}
\eeq
{}From the leading order result (\ref{eqn5.18}) we expect the local value of the 
Hubble parameter in the corresponding cosmological-like bubble to be
\beq
H_i \sim \sqrt{\dot{\phi}^2} \sim \vert N' (v_i) \vert \sim \frac{\vert N (v_i) 
\vert}{\ell_i}
\label{eq6.3}
\eeq
where $v_i$ denotes the advanced time at which the instability sets in. Combining 
(\ref{eq6.2}) and (\ref{eq6.3}) we get
\beq
H_i \, L_i \sim \vert N (v_i) \vert \gtrsim 1 \, . 
\label{eq6.4}
\eeq
These rough formulae show how, in principle,  given the stochastic 
properties of the dimensionless news functions, one could deduce the 
distribution of $H_i$ and $L_i$, naturally constrained by $H_i \, L_i \gtrsim 
1$. The corresponding distribution of $g_i$ is {\it a priori} independent from 
that of $\ell_i \sim L_i$, being linked to the presence of a slowly-varying (``DC") 
component $N_0$ 
in $N(v)$ (by contrast to the local variations of $N(v)$ leading to 
instability): $N(v) = N_0 + {\rm fluctuations}$. 
The DC component $N_0$ corresponds to ramps in $f(v)$, 
$f(v) = N_0 \, v + f_0 + {\rm fluctuations}$, i.e. to shifts of $\phi_i : \phi_i 
\rightarrow \phi_i + 2 N_0$. Finally, we can consider that the initial 
distribution of the news functions defines an {\it a priori} probability 
distribution for $g_i$, and $H_i$,
\beq
dp_{\rm prior} = w_i (g_i , L_i) \, dg_i \, dL_i \, ,
\label{eq6.5}
\eeq
with $H_i$ generically of order $L_i^{-1}$ (because $\vert N (v_i) \vert \sim 1$ 
is the threshold for instability).

 An important remark must be made here. The 
two basic parameters we are talking about, say $g_i$ and $L_i \sim H_i^{-1}$, 
precisely correspond to the two {\it global symmetries}, 
Eqs.~(\ref{p1})--(\ref{eqn2.9}) and Eq.~(\ref{eqn2.10}),
of the classical string 
vacua: a constant shift in $\phi$, and a global coordinate rescaling. If we consider that 
the initial state of string theory is classical (rather than quantum), these 
symmetries mean that no particular values for $\phi_i$ or $L_i$ are {\it a 
priori} preferred. One might even expect a ``flat'' distribution compatible with 
these global symmetries,
\beq
dp_{\rm prior}^{\rm classical \, flat} \propto \frac{dg_i}{g_i} \, 
\frac{dL_i}{L_i} \propto d\phi_i \, d \log L_i \, .
\label{eq6.6}
\eeq
Evidently, the problem with such a flat distribution is that it is non-normalizable.
 Some cut-offs are needed to make sense of such a flat prior 
distribution but, while there are natural  strong coupling/curvature  cut-offs
(the limits of validity of our approximation), it is not so easy to find
natural small-curvature/coupling cut-offs. If we appeal to string theory
 for providing cut-offs for $g_i$ and 
$L_i$, arguably the prior distribution (\ref{eq6.5}) should be determined by 
the conjectured basic symmetries of string theory, i.e. by $S$ and 
$T$-duality. $S$-duality suggests that one should work with the complex quantity 
$z = \frac{\theta}{2\pi} + i \, \frac{4\pi}{g^2}$ and require modular invariance 
in $z = x + iy$. This selects the Poincar\'e 
metric $ds^2 = (dx^2 + dy^2) / y^2$ 
as being special, and thereby defines a preferred measure in the fundamental 
domain (key-hole region) of the $(x,y)$ 
plane: $d\mu = \sqrt g \, dx \, dy = dx \, dy / y^2 = dx \, \vert dy^{-1} 
\vert$. Integrating out the angular variable $\theta$ over $0 \leq \theta < 
2\pi$ (which is the correct key-hole range when $g^2 < 4\pi$, i.e. $y > 1$) we 
end up with a preferred probability distribution for $g : dp(g) \propto dg^2 
\propto g \, dg$, considered in the weak-coupling region $g \lesssim 1$. A 
similar argument based on $T$-duality selects as preferred probability 
distribution for the spatial length scale $L : dp (L) \propto d 
\widehat{L}^{-2} \,  \propto 
\widehat{L}^{-3} \, d \widehat{L}$ where $\widehat{L} \equiv L / \ell_s$ is 
considered in the weak-$\sigma$-model-coupling region $\widehat{L} \gtrsim 1$. 
In conclusion, this type of argument would suggest the following normalizable, 
factorized prior distribution
\beq
dp_{\rm prior}^{\rm S-T-dual} \propto (g_i \, dg_i) \, \widehat{L}_i^{-3} \, d 
\widehat{L}_i \propto g_i \, dg_i \, \widehat{H}_i \, d \widehat{H}_i\,,
\label{eq6.7}
\eeq
with $0 < g_i \lesssim 1$, $0 < \widehat{H}_i \equiv H_i / M_s = 
\widehat{L}_i^{-1} \lesssim 1$. 

For the sake of 
generality, we wish  to leave open the nature of the prior distribution, 
and thus consider  the general class of prior probability distributions:
\beq
dp_{\rm prior} \propto g_i^a \, dg_i \, \widehat{H}_i^b \, d \widehat{H}_i\,,
\label{eq6.8}
\eeq
with arbitrary powers $a$ and $b$.

Any given initial distribution, say (\ref{eq6.8}), will generate a 
corresponding ensemble of PBB inflationary bubbles. We assume, as usual,
the existence of a successful ``exit'' mechanism by which 
dilaton-driven inflation, with growing $g$ and $H$  
finally ``exits'' into a standard hot Big Bang, i.e. a radiation-dominated 
Friedmann universe. Current ideas about how this might happen \cite{GMV} assume that 
the Hubble
parameter  reaches 
values of order of the string mass $M_s$, i.e. $\widehat{H}_s = H_s / M_s 
\sim 1$, {\it before} the string coupling constant reaches values of order 
unity. If the opposite were true ($g \sim 1$  occurring before 
$\widehat{H} \sim 1$), it is likely that quantum fluctuations would take
over before $H$ reaches the string scale, causing
 inhomogeneities to grow
so large that the inflationary process ``aborts" before a baby
universe is born. In any event, we shall discard this possibility, assuming that
the resulting cosmological universe would not evolve into anything able 
to harbour scientific civilizations.

After $\widehat{H}$ reaches $1$ at $g \ll 1$, 
 the evolving universe is expected to be entering a so-called stringy 
De~Sitter phase, i.e. a phase during which  $\dot{\phi}$ and 
$H$ are constant and given by some fixed-point values  of the order of $M_s$:
 $\vert \dot{\phi} \vert = x \, M_s$, and $\vert H \vert = y \, 
M_s$ with $x \sim y \sim 1$. [See \cite{GMV} for some suggestions on how to 
implement this mechanism.] Only the ratio between $x$ and $y$ enters our 
present phenomenological discussion. In keeping with a notation used in 
previous papers on PBB phenomenology, we introduce the (positive) parameter 
$\beta$ by
\beq
\vert (\ln g)^{\cdot} \vert = \frac{1}{2} \, \vert \dot{\phi} \vert = \beta \, H 
= {\rm const} \, .
\label{eq6.9}
\eeq

One finally assumes that this stringy De~Sitter-like phase 
ends when $g = e^{\phi/ 2}$ 
reaches values of order unity, causing the amplified vacuum 
fluctuations
to reach the critical density. After this moment, $g$ should remain 
fixed at about its present value, either because of a nonperturbative $\phi$-dependent potential, 
or possibly because of the attractor mechanism of Ref.~\cite{DP}. The fixing of 
$g$ marks the end of the stringy modifications to Einstein's theory, and the 
beginning of standard Friedmann-like cosmology. As we assume that $g$ is 
attracted towards a unique fixed point, the resulting Friedmann universe
will have no free parameter which could make it different
from our universe, except for its initial homogeneity scale $L_f$, 
where the index $f$ refers to the final state of the stringy phase. Therefore, 
the number of civilizations in the resulting Friedmann universe will be 
proportional to its total volume, at least for the large enough 
(in fact, 
old enough)  universes that can harbour life
\beq
{\cal N}_{\rm civ} (g_i , H_i) \propto {\cal V}_f (g_i , H_i) \, \theta [{\cal 
V}_f - {\cal V}_{\min}] \, .
\label{eq6.10}
\eeq
One assumes here, \`a la Dicke \cite{D61}, that the time span during which 
civilizations can occur, being constrained by the lifetime of stars, is fixed. 
The minimum scale ${\cal V}_{\rm min}$ for a life-harbouring universe is not known precisely, though 
it corresponds probably to universes whose total lifetime before recollapse is a 
few billion years. In conclusion, the {\it a posteriori} probability for a random 
scientific civilization to ask fine-tuning questions about the values of $g_i$ 
and $H_i$ is
\beq
dp_{\rm post} \propto {\cal N}_{\rm civ} (g_i , H_i) \, dp_{\rm prior} \propto 
{\cal V}_f (g_i , H_i) \, \theta ({\cal V}_f - {\cal V}_{\min}) \, dp_{\rm 
prior} \, .
\label{eq6.11}
\eeq
\subsection{Computation of the a posteriori probability distribution in 4 dimensions} 
\label{subsec6.3}
The computation of the final volume  ${\cal V}_f$ can be separated as
\beq
{\cal V}_f = L_f^3 = L_i^3 \ \frac{{\cal V}_1}{{\cal V}_i} \ z_s^3 \, .
\label{eq6.12}
\eeq
Here, the index $i$ refers to the (cosmological) time $t_i$ of beginning of 
dilaton-driven inflation, the index 1 refers to the time $t_1$ of transition between 
dilaton-driven superinflation and a De~Sitter-like phase, and $z_s = \exp \, (H 
\, (t_2 - t_1))$ denotes the total expansion during the De~Sitter phase. The 
ending $t_2$ of the De~Sitter-like phase is assumed to take place whence $g_2 
\equiv g (t_2) \sim {\cal O} (1)$.

Using the leading power-law behaviour of the cosmological evolution 
during the dilaton-driven phase, Eq.~(\ref{strkasn1}), 
and imposing that this phase ends when $H_1 
\sim (-t_1)^{-1} \sim M_s$, one gets  
\beq
\frac{{\cal V}_1}{{\cal V}_i} \simeq \left( \frac{t_i}{t_s} \right)^{-\Sigma 
\alpha} \simeq (\widehat{H}_i)^{\Sigma \alpha} \, ,
\label{eq6.13}
\eeq
where $\Sigma \alpha \equiv \alpha_1 + \alpha_2 + \alpha_3$ is the sum of the 
string-frame Kasner exponents. The $\alpha$'s are negative (expansion) and are 
constrained by Eq.~(\ref{strrel}). In the result (\ref{eq6.13}) we have neglected a factor 
of order unity linked to the fact that the dilaton-driven inflation is 
generically anisotropic ($\alpha_1 \not= \alpha_2 \not= \alpha_3$) and therefore 
that the various Hubble expansion rates will reach 
the string scale at slightly different times. We assume here that a nontrivial 
basin of attraction from anisotropic expansion toward an isotropic stringy 
De~Sitter-like phase exists as it is the case in examples \cite{GMV}.

During the subsequent stringy De~Sitter phase, the growth of $g$ is  related, according 
to Eq.~(\ref{eq6.9}), to the spatial expansion by $g \propto a^{\beta}$ 
so that
\beq
z_s = \frac{a_2}{a_1} = \left( \frac{g_2}{g_1} \right)^{\frac{1}{\beta}} \sim 
g_1^{-\frac{1}{\beta}} \, ,
\label{eq6.14}
\eeq
where we used $g_2 \sim 1$. On the other hand, $g_1 = 
g(t_1)$ is expressible, through the use of Eq.~(\ref{strkasn2}), giving the evolution of $g$ 
during the dilaton phase, as
\beq
e^{\phi_1 - \phi_i} = \left( \frac{g_1}{g_i} \right)^2 \simeq \left( 
\frac{t_1}{t_i} \right)^{\Sigma \alpha-1} \sim \widehat{H}_i^{\,\,\Sigma \alpha 
-1} \, .
\label{eq6.15}
\eeq
Finally, we express, using $L_i \sim H_i^{-1}$, the  volume at the beginning of 
standard cosmology in the form
\beq
{\cal V}_f (g_i , H_i) \sim \widehat{H}_i^{-A} \, [g_1 (g_i , 
\widehat{H}_i)]^{-3/\beta} \, \theta (1-g_1)\,,
\label{eq6.16}
\eeq
with 
\beq
g_1 (g_i , \widehat{H}_i) \sim g_i \, \widehat{H}_i^{-B} \,,
\label{eq6.17}
\eeq
and where the exponents $A$ and $B$ are positive, and are given in terms of the $\alpha$'s 
by
\beq
A = 3 - \Sigma \alpha \,, \quad \quad  
B = \frac{1}{2} - \frac{1}{2} \ \Sigma \alpha \, .
\label{eq6.18}
\eeq
If we assume, for illustration, an {\it a priori} distribution of values of $g_i$ and 
$\widehat{H}_i$ of the form (\ref{eq6.8}), we get for the {\it a posteriori} 
probability distribution of $g_i$ and $\widehat{H}_i$ 
\bea
dp_{\rm post} \propto g_i^a \, \widehat{H}_i^{b-A} \, [g_i \, 
\widehat{H}_i^{-B}]^{-3/\beta} \, \theta (1-g_1 (g_i , \widehat{H}_i)) \, \theta 
({\cal V}_f - {\cal V}_{\min}) \, dg_i \, d\widehat{H}_i \, ,
\eea
which can be written as
\beq
dp_{\rm post} \propto g_1^{\overline a} \, \widehat{H}_i^{-\overline A} \, 
\theta (1-g_1) \, \theta ({\cal V}_f - {\cal V}_{\min}) \, dg_1 \, d 
\widehat{H}_i \, .
\label{eq6.19}
\eeq
In the last form we have expressed the measure $dp_{\rm post}$ in terms of the 
two independent variables $g_1$ and $\widehat{H}_i$ 
(this is convenient because 
of the constraint $g_1 < 1$ carried by the step function). The exponents 
$\overline a$ and $\overline A$ appearing there are
\bea
&& \overline a \equiv a - \frac{3}{\beta} \,, \\
&& \overline A = A - (a+1) \, B-b = 3 + \frac{1}{2}\,(a-1)\,(\Sigma \alpha) - \frac{1}{2} \, 
(a+1) - b \, .
\label{eq6.20}
\eea
The numerical values of the exponents $\overline a$ and $\overline A$ play a 
crucial r\^ole in determining the {\it a posteriori} plausibility of our universe 
having evolved from the seemingly ``unnatural'' values (\ref{eq6.1}). Indeed, if 
$\overline A \geq 1$ the posterior probability distribution for $\widehat{H}_i$ 
is peaked, in a non integrable manner, at $\widehat{H}_i = 0$. Therefore, if 
$\overline A \geq 1$, it is natural to expect (within the present scenario) that 
the initial homogeneous patch, whose gravitational instability led to our 
universe, be extremely large compared to the string scale [and that the value of 
$g_i = \widehat{H}_i^B \, g_1$ with $B > 0$ and $g_1 < 1$ be correspondingly 
small]. Most scientific civilizations are bound to evolve in such universes.

Note that the duality-suggested values $a=1$ and $b=1$ of 
Eq.~(\ref{eq6.8}) lead to $\overline A = 1$ (independently of the $\alpha$'s) 
for which such an {\it a posteriori} explanation works. From this point of view one 
can argue that, within string theory, it is ``natural'' to observe very small 
numbers such as in Eq.~(\ref{eq6.1}). The exponent $\overline a$ also plays an 
important r\^ole. Indeed, if $\overline a > -1$, i.e. if
\beq
\beta > \frac{3}{a+1} \, ,
\label{eq6.21}
\eeq
the {\it a posteriori} probability distribution for $g_1$ is integrable over its 
entire possible range $(0,1)$. This means that most scientific civilizations are 
expected to observe values $g_1 = {\cal O} (1)$, corresponding to a small number 
of $e$-folds during the stringy phase: $z_s \sim g_1^{-1/\beta} = {\cal O} (1)$. 
This is a phenomenologically interesting case, because it means that various 
interesting physical phenomena taking place during the dilaton-driven phase 
(such as the quantum amplification of various fields) might leave observable 
imprints on cosmologically relevant scales. [A very long string phase would 
essentially iron out all signals coming from the dilaton phase.] It is 
interesting to note that, in the string-duality-inspired case (\ref{eq6.7}), 
the great divide between a short stringy phase $(\overline a > -1)$, and a very 
long one ($\overline a \leq -1$, implying that $g_1$ is peaked in a non 
integrable manner at $g_1 = 0$), lies at $\beta = 3/2$, which played already a 
special r\^ole in previous phenomenological studies \cite{PBB}.

Though the case $\overline A \geq 1$, $\overline a > -1$ appears as the 
conceptually most interesting case for the pre-Big Bang scenario, other 
civilization-related selection effects might also render the case 
$\overline A < 1$ viable. In this case, the factors 
$\widehat{H}_i^{-\overline A} \,d\,
\widehat{H}_i$ in Eq.~(\ref{eq6.19}) suggests that $\widehat{H}_i$ 
should be of order unity (i.e. $L_i \sim H_i^{-1} \sim \ell_s$). However, in 
such a case, one must take into account the factor $\theta ({\cal V}_f - {\cal 
V}_{\min})$ (which could be neglected in the previous discussion). This factor 
means that scientific civilizations should {\it a posteriori} expect to find 
themselves in the smallest possible universe compatible with their appearance. 
This probably means that they should also expect to see inhomogeneities 
comparable to the Hubble scale, and to have appeared very late in the 
cosmological evolution, just before the universe recollapses. This seems to conflict with 
our observations. In conclusion, within this model, and limiting our discussion 
for simplicity to the class (\ref{eq6.8}) with $a$ and $b$ integers, one finds 
the favorable situation $\overline A \geq 1$ realized (i.e. no {\it a posteriori} 
unnaturalness in observing very small $g_i$ and $\widehat{H}_i$) when either 
$a=0$ and $b \leq 2$, or $a=1$ and $b \leq 1$. However, if $a \geq 2$, i.e. if 
the {\it a priori} distribution function of $g_i$ vanishes faster than $g_i^2$ for 
$g_i \rightarrow 0$, the pre-Big Bang scenario has to face 
a naturalness problem. [This is the case for quantum fluctuations 
of $g_i$, as discussed at the end of the next subsection.]

\subsection{Computation of the a posteriori probability distribution in $D$ dimensions} 
\label{subsec6.4}
Assuming that the results discussed so far are qualitatively correct in 
higher dimensions,  
we have generalized the above considerations to the case in which the initial 
string vacuum is extended to $d$ spatial dimensions (say $d = 9$), and where, 
through gravitational instability, 3 of these dimensions are expanding and 6 are 
collapsing down to the string scale. This leads to introducing 4 special times: 
$t_i$ (beginning of dilaton phase), $t_1 \sim -t_s$ (end of the dilaton-driven 
power-law evolution of space), $t_2$ (when the radius of curvature of the 
collapsing dimensions become ${\cal O} (\ell_s)$), and $t_3$ 
(when $g_3 = {\cal O} (1)$). 
In this scenario there are two stringy De~Sitter-like phases: a first 
one during which the 3 ``external'' dimensions grow exponentially $a \propto 
\exp \, (H_1 \, t)$, while the $(d-3)$ ``internal'' ones {\it shrink} 
exponentially $\overline a \propto \exp \, (- \overline{H}_1 \, t)$ (and while 
$\dot{\phi} = 2\beta_1 \, H_1 = 2 \overline{\beta}_1 \, \overline{H}_1$), and a 
second phase during which the $(d-3)$ string-scale-curved internal dimensions have 
frozen while the 3 external ones continue to grow exponentially $a 
\propto \exp \, (H_2 \, t)$ (and while $\dot{\phi} = 2\beta_2 \, H_2$). This 
scenario gives a result of the same form as above, namely
\beq
\label{eq6.22}
{\cal V}_f (g_i , H_i) \sim \widehat{H}_i^{-A} \, 
\left [g_2(g_i,\widehat{H}_i) \right ]^{-3/\beta_2} \, \theta 
(1-g_2) \,,
\eeq
with  
\beq
g_2(g_i,\widehat{H}_i) = g_i\,\widehat{H}_i^{-B}\,, 
\eeq
and where
\beq
A = d - 3\alpha - (d-3) \, \overline{\alpha} + ( 1 - \overline{\alpha}) \, 
\overline{\beta}_1 \left( \frac{3}{\beta_1} - \frac{d-3}{\overline{\beta}_1} 
\right) \, ,
\eeq
\beq
B = \frac{1}{2} - \frac{3}{2} \, \alpha - \frac{1}{2} \, (d-3) \, 
\overline{\alpha} + (1 - \overline{\alpha}) \, \overline{\beta}_1 \, .
\label{eq6.23}
\eeq
Here one has assumed for simplicity that, during the dilaton phase, the 3 
external dimensions grow with the same exponent \footnote{
This string-frame Kasner exponent $\alpha$ should not be confused with the 
basic parameter $\alpha(v)$ introduced in Sec.~\ref{subsec5.3}.}
$\alpha < 0$, i.e. proportionally to 
$(-t)^{\alpha}$, while the $(d-3)$ internal ones decrease with the same exponent 
$\overline{\alpha} > 0$, proportionally to $(-t)^{\overline{\alpha}}$. Then, 
one can derive the analog of Eq.~(\ref{eq6.19}) with $g_1$ replaced by $g_2$ and 
with exponents
\beq
\overline a = a - \frac{3}{\beta_2} \, 
\quad \quad \overline A = A - (a+1) \, B - b \, .
\label{eq6.24}
\eeq
Once more, $\overline A \geq 1$ leads to no fine-tuning and $\overline a > -1$ to a 
short string phase. We shall not attempt a complete phenomenological discussion 
of the range of values of $a$ and $b$ compatible with $\overline A \geq 1$. 
(This would imply exploring the parameter space of allowed values for the Kasner 
exponents $\alpha$ and $\overline{\alpha}$.) Let us only note here that the 
introduction of more spatial dimensions generically helps because $A$ (and 
$\overline A$) now contains the contribution $+d = +9$, say, instead of 
the previous $+3$.
In this case, even  distribution functions vanishing faster than $g_i^2$ 
(for $g_i \rightarrow 0$) can make very small values of $g_i$ 
be {\it a posteriori} preferred. The power-law enhancement brought by the volume 
factor ${\cal V}_f$ would  be inadequate, however, for compensating  
an exponential suppression as $g_i \rightarrow 0$, 
say $\propto \exp \, (-c / g_i^2)$. In particular, the {\it a priori} distribution of $g_i$ 
and $H_i$ cannot be the one expected (from the Lagrangian ${\cal L} \sim 
e^{-\phi} (\nabla \phi)^2 \sim (\nabla g^{-1})^2$) for quantum fluctuations of 
$g$ on scales $L_i \sim H_i^{-1}$ around the trivial ground state. Indeed, 
this distribution vanishes 
for small $g_i$ and small $\widehat{H}_i$ as $\exp \, (-c\,g_i^{-2} \, 
\widehat{H}_i^{-2})$ with $c$ of order unity. 
This confirms the standard idea of the pre-Big Bang scenario, 
according to which 
the initial state should be a {\it classical} string vacuum, i.e. an 
arbitrary classical solution of the low-energy field equations.

\section{Conclusion}

We have proposed a new, concrete  picture for the 
typical initial state of the Universe which can give rise
to a cosmology of the so-called pre-big bang type. We conceived this state as a very 
generic, past-trivial string vacuum, i.e., classically, as the most general solution of 
the low-energy tree-level string action endowed with a perturbative region in the asymptotic 
past. Such states can be parametrized (in three spatial dimensions) by three 
dimensionless ``news functions'': a scalar news $N(v,\theta ,\varphi)$ and two 
helicity-2 news $N_+ (v,\theta ,\varphi)$ and $N_{\times} (v,\theta ,\varphi)$. 
By studying the simple spherically symmetric tensor-scalar model we have argued 
that such a generic ``in state'' will be gravitationally unstable if (any of) 
the news functions varie(s) by something of order unity on some scale $\Delta v 
\sim \ell$. More precisely, in the spherically symmetric model, we found, at 
lowest order of perturbation theory, that the criterion for gravitational 
instability is given by the supremum over $v_1$ and $v_2$ of the {\it variance} 
of the function $N(x)$ over the interval $[v_1 , v_2]$, see Eq.~(\ref{eqn4.16}). When this 
supremum exceeds a threshold of order unity, the wave packet ``sent'' between 
the retarded times $v_1$ and $v_2$ will become gravitationally unstable when, later, 
it becomes focussed at the center \footnote{In absence of spherical symmetry, 
there may exist many ``local centers'' where gravitational instability sets in. 
They should each correspond to an annular section of ${\cal I}^-$ with a ${\cal 
O} (1)$ fluctuation of the news looking roughly spherically symmetric when 
viewed from some center and in some suitably boosted Lorentz frame.}. We have 
verified that the infinite region between past infinity and such a center can be 
described by weak-field perturbation theory. Then, near the center, in a region 
of spatial extension $L_i \sim \Delta v \equiv v_2 - v_1$, one can abruptly 
shift from a weak-field description to a cosmological one. One can then describe 
the further evolution of the patch $L_i$ by a complementary cosmological-like 
expansion of the type introduced by Belinsky, Khalatnikov and Lifshitz (BKL)
and recently developed, for the PBB scenario, in \cite{V97}. We 
have proven to all orders of iteration (in our toy model) that the presence of 
the dilaton ensured the consistency of a (non oscillatory) BKL expansion in 
log-corrected powers of the distance to the future (space-like) singularity. This 
result confirms the smoothing behaviour of a dilaton-driven superinflationary 
evolution. In the string frame, gravitationally unstable patches of vacuum, with 
homogeneity scale $L_i$, will expand (if $\dot{\phi} > 0$) into homogeneous patches 
 on much larger scales. These blistering patches are surrounded by 
contracting regions where $\dot{\phi} < 0$. Many such dilaton-driven pre-Big 
Bang inflationary bubbles, looking like closed universes, can blister off a 
generic past-trivial string vacuum. By taking into account the selection effects 
linked to the presence of scientific civilizations, we have shown that it can 
appear {\it a posteriori} probable to observe that our universe comes from a very 
large fluctuation $L_i \gg \ell_s$, with a very small $g_i$. In particular, we 
pointed out that string-dualities suggest an {\it a priori} distribution function for 
$g_i$ and $L_i$ which is phenomenologically encouraging in predicting a short, 
intermediate  De~Sitter-like stringy phase, and the {\it a posteriori} naturalness of 
observing $L_i \gg \ell_s$ and $g_i \ll 1$.

Our scenario contains new features that might suggest new observable signatures 
of the pre-Big Bang idea. In particular, we expect the dilaton-driven phase 
to be generically anisotropic. It will be interesting to study the possible 
fossils of this anisotropy, especially in the presence of a short string phase. Our 
work suggests also new types of criteria for the gravitational instability of the 
Einstein-plus-scalar system. Concentrating on some non cumulative measure of the 
variation of the news might lead to new theorems of the type of those of 
Ref.~\cite{chi91}, \cite{chi93}.

One of the most interesting well-posed technical problems suggested by our work 
is the study of the link between the incoming news function and the spatial 
dependence (along the singularity) of the Kasner exponents. We hope to study 
this problem in the near future.

\acknowledgments

We thank Valodia Belinsky, Demetrios Christodoulou, Nemanja Kaloper, 
Sergiu Klainerman, Andrei Linde  and Slava 
Mukhanov for useful exchanges of ideas. Discussions with Amit Gosh, 
Federico Piazza, Giuseppe Pollifrone and Eliezer Rabinovici 
are also acknowledged. 

\appendix
\section{}
\label{app}
In this appendix we will establish the structure of 
the perturbative expansion near 
the singularity in the spherically symmetric model analysed in the text. 
We found it more convenient to use the $(v,r)$ coordinate system 
introduced in Sec.~\ref{sec3}. The field equations read
\beq
\label{a2}
2 \left [ \frac{\pa^2 \phi}{\pa v \pa r } + \frac{1}{r}\,
 \left (\frac{\pa \phi}{\pa v}\right ) \right ] + e^\beta \left (1-\frac{2m}{r} \right )\,
 \left ( \frac{\pa^2 \phi}{\pa r^2}\right )  +{ 2e^\beta \over r} \,
\left (1-\frac{m}{r} \right)
 \left (\frac{\pa \phi}{\pa r} \right ) = 0\,,
\eeq
\beq
\label{a1}
\left (\frac{\pa \beta}{\pa r}\right ) = \frac{r}{4}\,
\left (\frac{\pa \phi}{\pa r}\right )^2 \,, \quad \quad 
2\left (\frac{\pa m}{\pa r}\right ) = \left ( 1 - \frac{2 m}{r} \right )\,
\frac{r^2}{4}\,\left (\frac{\partial \phi}{\pa r} \right )^2 \,,
\eeq
Let us expand the different fields $(\phi, \beta, m)$ around 
the leading-order solutions (\ref{mass}), (\ref{phi}) and (\ref{beta})
\bea
\label{a11}
\phi_0(v,r) &=& C_{\phi}(v) + 2\alpha(v)\,\log r\,,\\
\label{a12}
\beta_0(v,r) &=& C_{\beta}(v) + \alpha^2(v)\,\log r\,,\\
\label{a13}
m_0(v,r) &=& C_{m}(v)\,r^{-\alpha^2(v)}\,,
\eea
and define the corrections to it $\ophi, \obeta$ and $\om$ by
\bea
\label{x1}
\phi(v,r) &=& \phi_0(v,r) + \ophi(v,r) \,,\\
\label{x2}
\beta(v,r) &=& \beta_0(v,r) + \obeta(v,r) \,,\\
\label{x3}
m(v,r) &=& m_0(v,r)\,(1+ \om(v,r)) \,.
\eea
Introducing the variable $ e^\lambda = r$, and denoting differentation
with respect to $v$ by a prime, 
Eqs.~(\ref{a2})--(\ref{a1}) take the form: 
\bea
\label{a5}
\frac{\pa^2 \ophi}{\pa \lambda^2} &=& S_{\phi}(v,\lambda) + 
L_{\phi}(v,\lambda) + {NL}_{\phi}(v,\lambda)\,, \\
S_{\phi} &=& \frac{e^{-C_\beta}}{2C_m}\,\left [
4\alpha^\prime\,(1 + \lambda)\,e^{2 \lambda} + 
2C_\phi^\prime\,e^{2 \lambda} + 2 \alpha\,e^{C_\beta}\,
e^{(\alpha^2+1)\,\lambda} \right ]\,,\\
L_{\phi} &=& \frac{e^{-C_\beta}}{C_m}\,\left [ 
e^{2 \lambda}\,\frac{\pa \ophi^\prime}{\pa \lambda}  +
e^{2 \lambda}\,{\ophi^\prime} + 
\alpha\,\obeta\,e^{C_\beta}\,e^{(\alpha^2+1)\,\lambda} +
\frac{e^{C_\beta}}{2}\,e^{(\alpha^2+1)\,\lambda}\,
\left ( \frac{\pa^2 \ophi}{\pa \lambda^2} + 
\frac{\pa \ophi}{\pa \lambda} \right ) \right ]\,,\\
{NL}_{\phi} &=& \frac{1}{2C_m}\,\left [
e^{(\alpha^2+1)\,\lambda}\,(e^{\obeta} -1)\, 
\left ( \frac{\pa^2 \ophi}{\pa \lambda^2} + 
\frac{\pa \ophi}{\pa \lambda} \right ) 
- 2C_m\, (e^{\obeta}-1)\,(1+\om)\,  \frac{\pa^2 \ophi}{\pa \lambda^2}\, +
\right . \nonumber \\
&& \left . 2 \alpha\,e^{(\alpha^2+1)\,\lambda}\,(e^{\obeta}-1-\obeta) 
-2C_m\,\om\,\frac{\pa^2 \ophi}{\pa \lambda^2} \right ]\,;
\eea
\bea
\label{a3}
&& \left (\frac{\pa \obeta}{\pa \lambda}\right ) - 
\alpha\,\left (\frac{\pa \ophi}{\pa \lambda}\right ) = 
S_\beta(v, \lambda) + L_\beta(v, \lambda) + {NL}_{\phi}(v, \lambda) \,,\\
&& S_\beta = 0= L_\beta \,, \quad \quad {NL}_{\phi} = 
\frac{1}{4}\,\left (\frac{\pa \ophi}{\pa \lambda}\right )^2\,;
\eea
\bea
\label{a4}
&& \left (\frac{\pa \om}{\pa \lambda}\right ) + 
\alpha\,\left (\frac{\pa \ophi}{\pa \lambda}\right ) = 
S_m(v,\lambda) + L_m(v,\lambda) + {NL}_m(v,\lambda)\,, \\
&& S_m = \frac{\alpha^2\,e^{(\alpha^2+1)\,\lambda}}{2C_m}\,, 
\quad \quad 
L_m = \frac{\alpha\,e^{(\alpha^2+1)\,\lambda}}{2C_m}\,
\left (\frac{\pa \ophi}{\pa \lambda}\right )\,,\\
&& {NL}_m = \frac{e^{(\alpha^2 + 1)\,\lambda}}{8C_m}\,
\left (\frac{\pa \ophi}{\pa \lambda}\right )^2 - 
\frac{1}{4}\,\left (\frac{\pa \ophi}{\pa \lambda}\right )^2
-\alpha\,\om\,\left (\frac{\pa \ophi}{\pa \lambda}\right ) 
- \frac{\om}{4}\, \left (\frac{\pa \ophi}{\pa \lambda}\right )^2\,.
\eea
In the R.H.S.'s of Eqs.~(\ref{a5}), (\ref{a3}) and (\ref{a4}) 
we have indicated by $S$ the field-independent terms (``source terms''), 
by $L$ the terms that are 
linear in the fields, but subleading  with respect to the L.H.S. 
in the limit $ r \rightarrow 0$ ($\lambda \rightarrow - \infty$), while 
by $NL$ we refer to generic non-linear contributions. 

By examining the source terms and the subleading linear terms, 
we see that they all contain either an explicit factor
$e^{(\alpha^2+1)\,\lambda} \equiv r^{(\alpha^2+1)}$ or 
a factor $e^{2 \lambda} \equiv r^2$. To keep 
track of the presence of these factors (which tend to zero 
at the singularity) let us introduce two formal bookkeeping 
parameters: $\sigma_1$ associated with each occurrence 
of $e^{(\alpha^2+1)\,\lambda}$ in $S$ and $L$ and, likewise, 
$\sigma_2$ associated with $e^{2\lambda}$. In the end we shall set 
$\sigma_1 = \sigma_2=1$, but the introduction of these parameters 
will be useful to delineate the structure of the expansion of 
the fields near $r = 0$. 

Introducing the notation $\psi = (\ophi, \obeta, \om)$, considered 
as a column matrix, and
denoting $D_\lambda$ the following derivative operator with respect to $\lambda$,  
\beq
D_\lambda = 
\left (
\begin{array}{ccc}
\pa_\lambda^2 & 0 & 0 \\
-\alpha(v)\pa_\lambda & \pa_\lambda & 0 \\
 + \alpha(v)\pa_\lambda & 0 & \pa_\lambda 
\end{array}
\right )\,,
\eeq
we can rewrite Eqs.~(\ref{a5}), (\ref{a3}) and (\ref{a4}) 
in the compact matrix form: 
\beq
\label{a18}
D_\lambda \psi = \sigma_1\,S_{\psi}^{(1)} + \sigma_2\,S_{\psi}^{(2)} + 
\sigma_1\,L_{\psi}^{(1)} + \sigma_2\,L_{\psi}^{(2)} + {NL}_{\psi}\,.
\eeq  
The nonlinear terms in (\ref{a18}) depend linearly on $\sigma_1$ 
and $\sigma_2$: ${NL}_\psi(\psi, \pa_\lambda \psi,\pa_\lambda^2 \psi, 
\sigma_1, \sigma_2) = {NL}_\psi^{(0)} + \sigma_1\,{NL}_\psi^{(1)} + 
\sigma_2\,{NL}_\psi^{(2)}$. The differential matrix system (\ref{a18})
can be rewritten as an integro-differential system by 
introducing the ``retarded'' Green function $G(\lambda, \lambda^\prime)$, 
i.e. the unique inverse of $D_\lambda$ which vanishes when $\lambda^\prime 
> \lambda$
\beq
\label{a19}
\psi = G * \left (\sigma_1\,S_{\psi}^{(1)} + \sigma_2\,S_{\psi}^{(2)} + 
\sigma_1\,L_{\psi}^{(1)} + \sigma_2\,L_{\psi}^{(2)} + {NL}_{\psi}\right )\,,
\eeq
where we indicate by $*$ the operation of integrating over $\lambda^\prime$.
This choice of boundary conditions is imposed by consistency:
We defined the background solution (\ref{a11})--(\ref{a13}) 
as giving the asymptotic behaviour of $(\phi, \beta, m)$ 
when $ r \rightarrow 0$, i.e. $\lambda \rightarrow -\infty$  . 
We must therefore require that the deviation field $\psi(\lambda)$ 
vanish when  $\lambda \rightarrow -\infty$.

The system (\ref{a19}) can be formally solved by successive 
iterations, leading to a solution $\psi$ in the form of a double 
power series in the bookkeeping parameters $\sigma_1$ and $\sigma_2$:
\beq
\psi = \sum_{n,m \geq 1} \sigma_1^n\,\sigma_2^m\,\psi_{n m}\,.
\label{Aexp}
\eeq
For instance, the first order (in $\sigma_1$ and $\sigma_2$) 
corrections 
$\psi_1 = \sigma_1\,\psi_{1 0} + \sigma_2\,\psi_{0 1}$, i.e. 
the next-to-leading-order terms in Eqs.~(\ref{x1})--(\ref{x3}), are obtained
by neglecting the $L$ and $NL$ terms. They satisfy the equations
\bea
\label{a7}
&& \frac{\pa^2 \phi_1}{\pa \lambda^2} = S_{\phi}(v,\lambda)\,, \\
\label{a6}
&& \left (\frac{\pa \beta_1}{\pa \lambda}\right ) - 
\alpha(v) \left (\frac{\pa \phi_1}{\pa \lambda}\right ) = 
S_\beta(v,\lambda)\,, \\
\label{a8}
&& \left (\frac{\pa m_1}{\pa \lambda}\right ) + 
\alpha(v) \left (\frac{\pa \phi_1}{\pa \lambda}\right ) = S_m(v,\lambda)\,.
\eea
The unique solution of this system which vanishes when 
$\lambda \rightarrow - \infty$ reads
\bea 
\label{a22}
\phi_1(v,\lambda) &=& \int^\lambda_{-\infty} d \lambda^\prime \,
\int^{\lambda^\prime}_{-\infty}d \lambda^{\prime \prime}\,  
S_\phi(v,\lambda^{\prime \prime }) = 
\int^\lambda_{-\infty} d \lambda^\prime \,
(\lambda -\lambda^\prime)\, 
S_\phi(v,\lambda^{\prime})\,, \\
\label{a21}
\beta_1(v,\lambda) &=& \alpha(v)\,\phi_1(v,\lambda) 
+ \int^\lambda_{-\infty} d \lambda^\prime \,
S_\beta(v,\lambda^{\prime})\,, \\
\label{a23}
m_1(v,\lambda) &=& -\alpha(v)\,\phi_1(v,\lambda)  + 
\int^\lambda_{-\infty} d \lambda^\prime\,S_m(v,\lambda^\prime)\,.
\eea
The form of this solution defines the action 
$G * S \equiv \int d \lambda^\prime \,G(\lambda, \lambda^\prime)\,
S(\lambda^\prime)$
of the retarded matrix Green function $G(\lambda, \lambda^\prime)$ 
on any source term $S(\lambda^\prime) = (S_\phi(\lambda^\prime), 
S_\beta(\lambda^\prime), S_m(\lambda^\prime))$. 
As a check on the consistency 
of the iteration method, note that, if one had not imposed any boundary 
conditions, the generic solution $\psi_1$ would have contained 
an arbitrary homogeneous solution $\psi_{\rm hom}$ 
(``zero mode''). However, it is easy to check 
that the addition of such a generic $\psi_{\rm hom}$ would be
absorbable in suitable redefinitions of the ``seed'' functions 
$\alpha(v), C_\phi(v)$, $C_\beta(v)$, $C_m(v)$ entering the background 
solution Eqs.~(\ref{a11})--(\ref{a13}). 
Finally, the explicit form of the next-to-leading 
order solution (with $\sigma_1 = \sigma_2 =1$) is 
\bea
\phi_1(v,\lambda) &=& \frac{\alpha}{(\alpha^2 +1)^2\,C_m}\,{e^{(\alpha^2+1)\,\lambda}}\,
+ \frac{C_{\phi}^\prime\,e^{-C_\beta}}{4C_m}\,e^{2 \lambda} + 
\frac{\alpha^\prime\,e^{-C_\beta}}{2C_m}\,\lambda\,e^{2 \lambda}\,, \\
\vrsmall
\beta_1(v,\lambda) &=& \alpha\,\phi_1(v,\lambda)\,,\\
m_1(v,\lambda) &=& -\alpha\,\phi_1(v,\lambda) + \frac{\alpha^2}{2(\alpha^2+1)C_m}\,
{e^{(\alpha^2+1)\,\lambda}}\,.
\eea
Note that the coefficients of the terms proportional to 
$e^{2 \lambda}$ and $\lambda\,e^{2 \lambda}$ contain derivatives 
with respect to $v$, hence they determine the inhomogeneous 
expansion near the singularity, 
while the terms proportional to $e^{(\alpha^2+1)\,\lambda}$ 
are present also in the purely homogeneous case.

Inserting the double power series expansion (\ref{Aexp}) 
into (\ref{a18}) or (\ref{a19}) yields, for each contribution $\psi_{n m}$, 
\beq
D_\lambda \psi_{n m} = s_{n m}\,, \quad \quad
\psi_{n m} = G * s_{n m}\,,
\label{anm}
\eeq
where the R.H.S. $s_{n m}$ plays, at each order, the r\^ole of an {\it effective 
source term} which  is (in principle) known, being 
some algebraico-differential expression in the lower-order 
terms $\psi_{n^\prime m^\prime}$, with $n^\prime + m^\prime < n+m$. 
Let us make the inductive assumption that $s_{n m}$ 
is a sum of terms of the form 
\beq
s_{n m l} = e^{n\,(\alpha^2+1)\,\lambda}\, e^{2 m\,\lambda}\, Q_l(\lambda)\,,
\label{a32}
\eeq
with $l \leq m $, where $Q_l(\lambda)$ is a polynomial in $\lambda$ 
of order $l$. [Here, $n, m, l$ are all integers].
It is straightforward to check, from the explicit action 
(\ref{a22})--(\ref{a23}) of the Green function on source terms, that the solution 
$\psi_{n m l}$ corresponding to any term $s_{n m l}$, Eq.~(\ref{a32}),  
in the R.H.S. of Eq.~(\ref{anm}) is again of the form 
\beq
\psi_{n ml} = G * s_{n m l} = 
e^{n\,(\alpha^2+1)\,\lambda}\, e^{2 m\,\lambda}\, \widetilde{Q}_l(\lambda)\,,
\eeq
where $\widetilde{Q}_l(\lambda)$ is another polynomial in $\lambda$, 
of the same order $l$. Examining the explicit form of the $L$ and $NL$ 
terms on the R.H.S. of Eqs.~(\ref{a5}), (\ref{a3}) and (\ref{a4}), 
one can then prove that the induction hypothesis consistently 
propagates at the next order in $n+m$ (still with the constraint 
$l \leq m $).

In conclusion, coming back to the variable $r$, we find that the 
inhomogeneous expansion near the singularity contains two intertwined 
series: an expansion in powers of $r^{(\alpha^2+1)}$ and one in 
powers of $r^2$ mixed with powers of $\log r$. The solution to all orders will be of the form: 
\beq
\psi = \sum_{n,\,m,\,l \leq m} a_{n m l}\,r^{n(\alpha^2 +1)}\,r^{2 m}\,P_l(\log r)\,,
\label{ulapp}
\eeq
where $a_{n m l}$, and the coefficients of the polynomial 
$P_l$, are 
functions of $v$ that in principle can be iteratively evaluated.
As shown in Sec.~\ref{subsec5.4}, the full subseries $\sum_n a_{n00}\,
r^{n\,(\alpha^2+1)}$ can be exactly summed in terms of the variable 
$x$ ($x \rightarrow 0^+$) such that $r^2 = l^2\,x^{1 +b}\,(1-x)^{1-b}$ 
with $b = (1-\alpha^2)/(1 + \alpha^2)$ (see Eqs.~(\ref{ap1}), (\ref{ap2})). On the other 
hand, the terms with $m \neq 0$ are more complicated, as 
they contain more and more $v$-derivatives of $\alpha(v)$, 
$C_\phi(v)$ and $C_\beta(v)$ as $m$ increases. 
We note, in passing, that this monotonous 
increase of  $v$-derivatives shows that the series 
in $m$ (and $l$) (inhomogeneous expansion) cannot
be convergent for generic (non analytic) seed functions $\alpha(v)$, 
$C_\phi(v)$ and $C_\beta(v)$. 

To complete the proof that the algorithm defining $\psi$ solves 
all the initial field equations (\ref{eq1.18})--(\ref{eq1.20}), we finally 
need to check that the $\om(v,r)$ constructed above satisfies 
the constraint (\ref{new}) involving $\pa m/ \pa v$. 
Let us denote the quantity which must vanish as 
\beq
{\cal E} \equiv \frac{1}{m} \left \{ 2 \left (\frac{\pa m}{\pa v}\right )_r - 
e^{-\beta}\,\frac{r^2}{2}\,\left [ \left (\frac{\pa \phi}{\pa v}\right )_r^2 +
e^\beta\,\left ( 1 - \frac{2 m}{r} \right )\, 
\left (\frac{\pa \phi}{\pa r}\right )_v\,
\left (\frac{\pa \phi}{\pa v}\right )_r \right ] \right \}\,.
\eeq
It is easily checked that, when the other field equations 
(\ref{eq1.18}), (\ref{eq1.19}) and (\ref{eq1.20}) are satisfied, 
${\cal E}$ fulfills the  following identity (which could also 
be derived from the Bianchi identities) 
\beq
\frac{\pa {\cal E}}{\pa r} + A(v,r)\,{\cal E}\equiv 0\,, 
\quad  \quad A(v,r) = \frac{r^2}{8\,m}\,\left ( \frac{\pa \phi}{\pa r} \right )^2\,.
\label{id}
\eeq
Noting that the coefficient $A(v,r)$ is bounded near $r=0$ ($A(v,r) \sim r^{\alpha^2(v)}$),
and that the relation (\ref{5.26}) ensures the vanishing of 
${\cal E}$ as $r \rightarrow 0$, we conclude from the identity (\ref{id}) 
(viewed as an ODE in $r$ with data given at $r=0$) that ${\cal E}$ 
vanishes everywhere.

\end{document}